\pdfoutput=1
\documentclass{article}
\usepackage{caption}
\usepackage{subcaption}
\usepackage{graphicx}
\usepackage{amsmath}
\usepackage{geometry}
\usepackage{authblk}
\usepackage{hyperref}
\usepackage[T1]{fontenc}
\usepackage[utf8]{inputenc}
\usepackage{float}
\usepackage{csquotes}
\usepackage[version=4]{mhchem}
\usepackage{amssymb}
\usepackage{authblk}
\usepackage{xcolor}
\usepackage{bbold}

\title{
Møller-Plesset Perturbation Theory Calculations on Quantum Devices
}
\author[1]{Junxu Li\thanks{Emial: lijunxu1996@gmail.com}}
\author[3]{Xingyu Gao}
\author[2]{Manas Sajjan}
\author[4]{Ji-Hu Su} 
\author[4]{Zhao-Kai Li}
\author[2, 3]{Sabre Kais\thanks{Email: kais@purdue.edu}}
\affil[1]{
Department of Physics,
College of Science,

Northeastern University,
Shenyang 110819, China
}
\affil[2]{Department of Chemistry,  and

Purdue Quantum Science and Engineering Institute

Purdue University, West Lafayette, IN 47907, United States}
\affil[3]{
Department of Physics and Astronomy,

Purdue University, West Lafayette, IN 47907, United States
}
\affil[4]{
Department of Modern Physics, University of Science and Technology of China, 

Hefei, Anhui, 230026, China}
\begin{document}

\maketitle

\begin{abstract}
Accurate electronic structure calculations might be one of the most anticipated applications of quantum computing.
% promising suggested applications of quantum computing.
The recent landscape of quantum simulations of chemistry within the Hartree-Fock approximation raises the prospect of substantial theory and hardware developments in this context.
%more intricate methods fulfilled on quantum devices.
% modeling the isomerization mechanism of diazene.
%In 2020, researchers from Google successfully simulated the HF wave function for hydrogen chains up to 12 qubits
Here we propose a general quantum circuit for Møller-Plesset perturbation theory (MPPT) calculations, which is a popular and powerful post-Hartree-Fock method widly harnessed in solving electronic structure problems.
MPPT improves on the Hartree–Fock method by including electron correlation effects wherewith Rayleigh–Schrödinger perturbation theory.
%As a typical application of Rayleigh–Schrödinger perturbation theory, in MPPT the Hartree-Fock Hamiltonian is regard as the unperturbed Hamiltonian, whereas the perturbation is the difference between Hartree-Fock Hamiltonian and real one.
%Given the energy levels and wavefunctions of Hartree-Fock approximation, 
Given the Hartree-Fock results, the proposed circuit is designed to estimate the second order energy corrections with MPPT methods.
In addition to demonstration of the theoretical scheme, the proposed circuit is further employed to calculate the second order energy correction for the ground state of Helium atom ($\ce{{}^4He}$), and the total error rate is around $2.3\%$.
Experiments on IBM 27-qubit quantum computers express the feasibility on near term quantum devices, and the capability to estimate the  second order energy correction accurately.
In imitation of the classical MPPT, our approach is non-heuristic, guaranteeing that all parameters in the circuit are directly determined by the given Hartree-Fock results.
Moreover, the proposed circuit shows a potential quantum speedup comparing to the traditional MPPT calculations.
Our work paves the way forward the implementation of more intricate post–Hartree–Fock methods on quantum hardware, enriching the toolkit solving electronic structure problems on quantum computing platforms. 
\end{abstract}

%%% MAIN %%%
\section*{Main}
%%% INTRO %%%
Recent landmarks in quantum hardware developments\cite{arute2019quantum, zhong2020quantum, kim2023evidence, king2023quantum, altman2021quantum}, along with innovations in developing quantum algorithms \cite{bravyi2018quantum, biamonte2017quantum, boixo2018characterizing, cong2019quantum, sajjan2022quantum}, herald the age of `quantum supremacy' dawns.
In this context, solving the classically intractable electronic structure problem might be one of the most promising applications of quantum computing\cite{aspuru2005simulated, wang2008quantum, lanyon2010towards, daskin2011decomposition, kais2014introduction, xia2018quantum, bian2019quantum, mcardle2020quantum, preskill2018quantum, mi2022time, lee2023evaluating}.
In the past decade, the variational quantum eigensolver (VQE)\cite{peruzzo2014variational, mcclean2016theory} has been widly employed in the electronic structure calculations\cite{kandala2017hardware, xia2020qubit, grimsley2019adaptive, huggins2021efficient}.
Extraordinarily in 2020, Google AI Quantum successfully implemented simulations of chemistry within the Hartree-Fock approximation for system sizes up to 12 qubits, which till now retains the record for the largest VQE calculations of ground state on quantum devices\cite{google2020hartree}.

The state-of-art quantum simulation within the Hartree-Fock approximation lays the foundation stone to implement more intricate {\it{ab initio}} methods on quantum hardware. 
Herein, we propose a general quantum circuit for Møller-Plesset perturbation theory (MPPT) calculations\cite{moller1934note}.
%%% OUTLINE %%%
To demonstrate the feasibility, the proposed circuit is employed to estimate the second order MP (MP2) correlation energy for the ground state of Helium atom ($\ce{{}^4He}$).
Experiments are carried on IBM 27-qubit quantum computers.

As a typical application of Rayleigh–Schrödinger perturbation theory\cite{schrodinger1926quantisierung}, in MPPT the Hartree-Fock Hamiltonian\cite{hartree1928wave, fock1930naherungsmethode} is regard as the unperturbed Hamiltonian, whereas the perturbation is the difference between Hartree-Fock Hamiltonian and the real one.
The MP2 correlation energy for ground state is\cite{szabo2012modern}
\begin{equation}
    E_0^{(2)} = \sum_{a<b}\sum_{r<s} 
    \frac{|\langle ab||rs\rangle|^2}{\epsilon_a+\epsilon_b-\epsilon_r-\epsilon_s}
    \label{eq_mp2}
\end{equation}
where $a,b$ indicate the occupied orbitals, $r,s$ indicate the virtual orbitals, $\epsilon$ is the orbital energies obtained from Hartree–Fock calculations, and $\langle ab||rs\rangle$ is the antisymmetrized two electron integral as shown in Eq.(\ref{eqm_anti}).
Here we focus on the ground state of Helium atom ($\ce{{}^4He}$), and the two occupied orbitals are both $1s$ orbitals, ensuring that
\begin{equation}
    E_0^{(2)} = \sum_{abrs}^{N/2}
    \frac{\langle ab|rs\rangle\langle rs|ab\rangle}
    {\epsilon_a+\epsilon_b-\epsilon_r-\epsilon_s}
\end{equation}
where $\langle ab|rs\rangle$ is the electron repulsion integral (ERI) under physicists notation, as shown in Eq.(\ref{eqm_eri_mo}).

%%% FIG MAIN %%%
\begin{figure}[htbp]
    \begin{center}
        \includegraphics[width=0.95\textwidth]{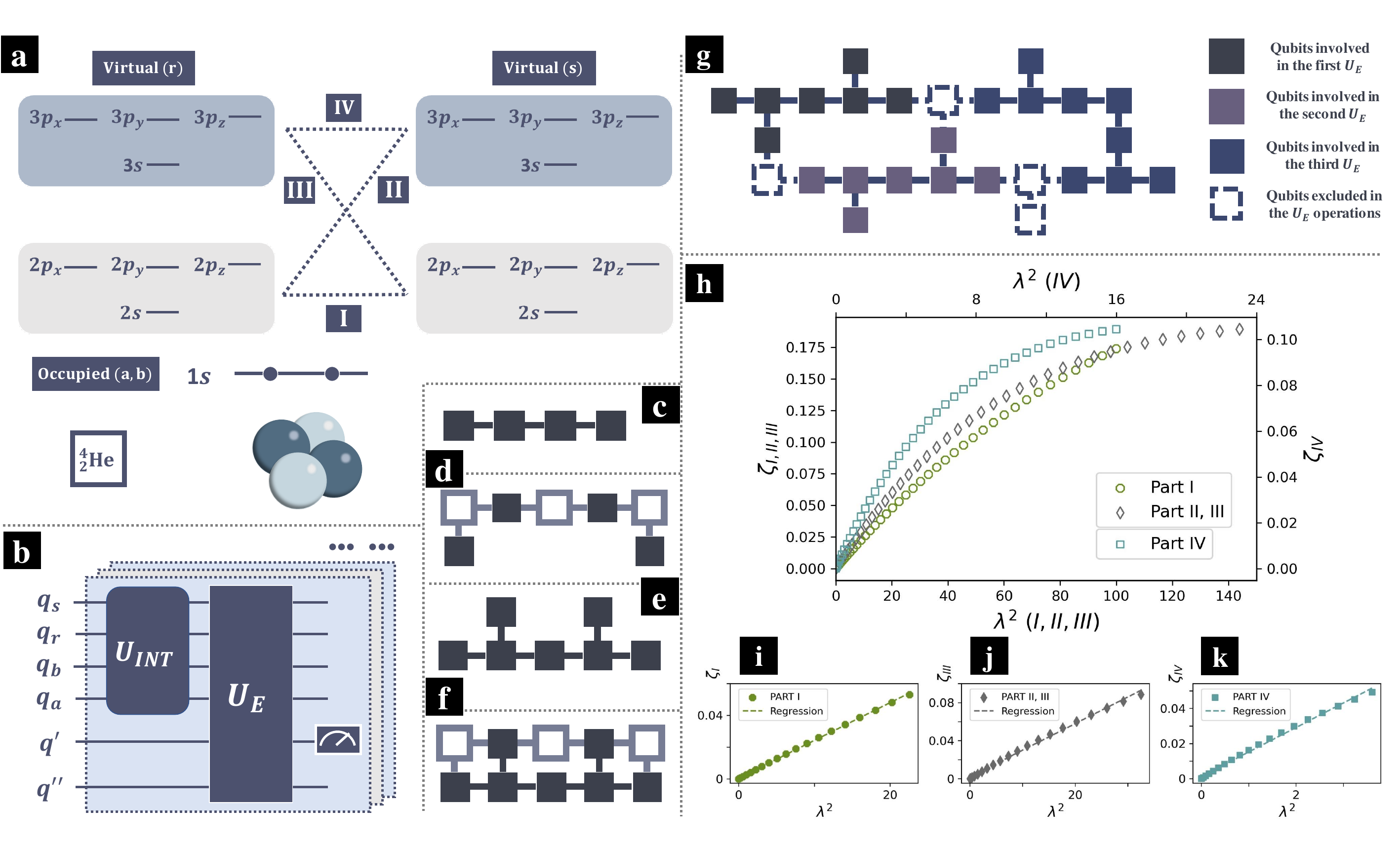}
    \end{center}
    \caption{
    {\bf Schematic depiction of MPPT methods and quantum circuit structures.}
    (a) Schematic depiction of MPPT methods.
    (b) Quantum circuit structures for the MPPT calculations, where $q_{a,b,r,s}$ correspond to the orbitals, $q'$ is the readout qubit and $q''$ are included as ancilla qubits.
    }
    (c) Necessary connectivity in the implementation of $U_{INT}$.
    In (c-g) Each square represents a qubit, and connected squares indicate connected ones on real device.
    (d) Connectivity in the implementation of $U_{INT}$, three ancilla qubits added (depicted as hollow squares).
    (e) Necessary connectivity in the implementation of $U_{E}$.
    (f) Necessary connectivity in the implementation of $U_{INT}$ together with $U_{E}$.
    (g) Schematic connectivity of IBM 27 qubit computer `ibm\_aucland', on which three $U_{E}$ can be tested in parallel, as depicted in the solid squares with various colors.
    (h-j) Simulation results of the $\zeta$ values with various $\lambda$.
    $\zeta$ is the probability to find $q'$ at state $|1\rangle$, where the subscripts indicate the four parts.
    \label{fig_main}
\end{figure}

The Hartree-Fock results such as molecular orbitals (MO), electronic repulsion integrals (ERI) are all obtained from PSI4\cite{parrish2017psi4}.
In the Hartree-Fock calculations, we use the basis `aug-cc-pvdz'\cite{kendall1992electron}, including $1s2s2p3s3p$ orbitals.
As depicted in Fig.(\ref{fig_main}a), there are only two electrons in Helium($\ce{{}^4He}$), occupying orbital $1s$ at ground state.
The virtual orbitals correspond to $2s2p3s3p$ orbitals, expanding the ERI tensor as a $16\times 16$ matrix, corresponding to indices $r,s$.
For simplicity, we divide the calculation into four parts, regarding to integrals between orbitals $2s2p-2s2p$(I), $2s2p-3s3p$(II), $3s3p-2s2p$(III) and $3s3p-3s3p$(IV).
These four parts can be calculated in parallel, and the schematic circuit implementation is presented in Fig.(\ref{fig_main}b), where $q_{a,b,r,s}$ correspond to the orbitals, $q'$ is the readout qubit and $q''$ are included as ancilla qubits.

There are two main operations, $U_E$ and $U_{INT}$, in the proposed circuit as shown in Fig.(\ref{fig_main}b).
In brief, operation $U_E$ generates the denominators $1/{(\epsilon_a+\epsilon_b-\epsilon_r-\epsilon_s)}$, whereas $U_{INT}$ prepares the ERIs.
$U_E$ is constructed with mainly multi-controller gates (See Eq.(\ref{eqm_ue})), which is designed to prepare accurate denominator terms.
On the contrary, $U_{INT}$ is designed to prepare approximations of the ERIs.
The ERIs matrix, or part of the ERIs, generally could not be mapped as a unitary operation directly.
Thereby, $U_{INT}$ is alternatively designed as $exp(i\lambda V)$, where $V$  corresponds to the ERIs, and $0<\lambda$.
$U_{INT}$ is implemented with the first order Trotter decomposition as shown in Eq.(\ref{eqm_uint}).
Here we focus on the ground state of Helium atom, qubits $q_{a,b}$ representing the occupied orbitals $a, b$ can be excluded in the implementation on hardware.

To begin with, all qubits are initialized at ground state $|0\rangle$.
Next, $U_{INT}(\lambda)$ converts $q_{a,b,r,s}$ into a certain state approximating the ERIs (or part of the ERIs).
$U_E$ then includes the denominator terms.
At the end the single qubit $q'$ is measured.
By repeating the process above with various $\lambda$ values, we are able to estimate the MP2 correlation energy. 
Denote $\zeta_{I,II,III,IV}$ as the probability to find $q'$ at state $|1\rangle$, where the subscripts indicate the four parts.
Theoretically, for small $\lambda$ values we have
\begin{equation}
    \left|E_0^{(2)}\right|
    = \frac{d}{d\lambda^2}(C_e^I\zeta_I+C_e^{II}\zeta_{II}+C_e^{III}\zeta_{III}+C_e^{IV}\zeta_{IV})+\mathcal{O}(\lambda)
    \label{eq_zeta}
\end{equation}
where $C_e$ is the constant in Eq.(\ref{eqm_ue}) ensuring $C_e/(\epsilon_a+\epsilon_b-\epsilon_r-\epsilon_s)\in[0, 1]$.
Superscripts of $C_e$ indicate the corresponding part.
In Fig.(\ref{fig_main}h-k) we present the simulation results of the $\zeta$ values with various $\lambda$.
As expected, $\zeta$ increases almost linearly with $\lambda^2$ for small $\lambda$ values, see Fig.(\ref{fig_main}i,j,k).
The linear trend changes around $\lambda=2$ for Part IV, and around $\lambda=5$ for other parts, due to the greater ERI components in Part IV.
Meanwhile, for the symmetry in ERIs and denominators, part II and part III make same contribution to the final result.
Thereby, we only need to estimate $\zeta_{I}$, $\zeta_{III}$ and $\zeta_{IV}$ to estimate the MP2 correlation energy $E_0^{(2)}$.

%%% RESULTS and DISCUSSION %%%
\section*{Results and Discussion}
In addition to numerical simulations, the proposed circuit is then employed on IBM 27-qubit quantum computer to calculate the MP2 energy correction for ground state of Helium atom ($\ce{{}^4He}$).

%%% FIG UE %%%
\begin{figure}[htbp]
    \centering
    \includegraphics[width=1.0\textwidth]{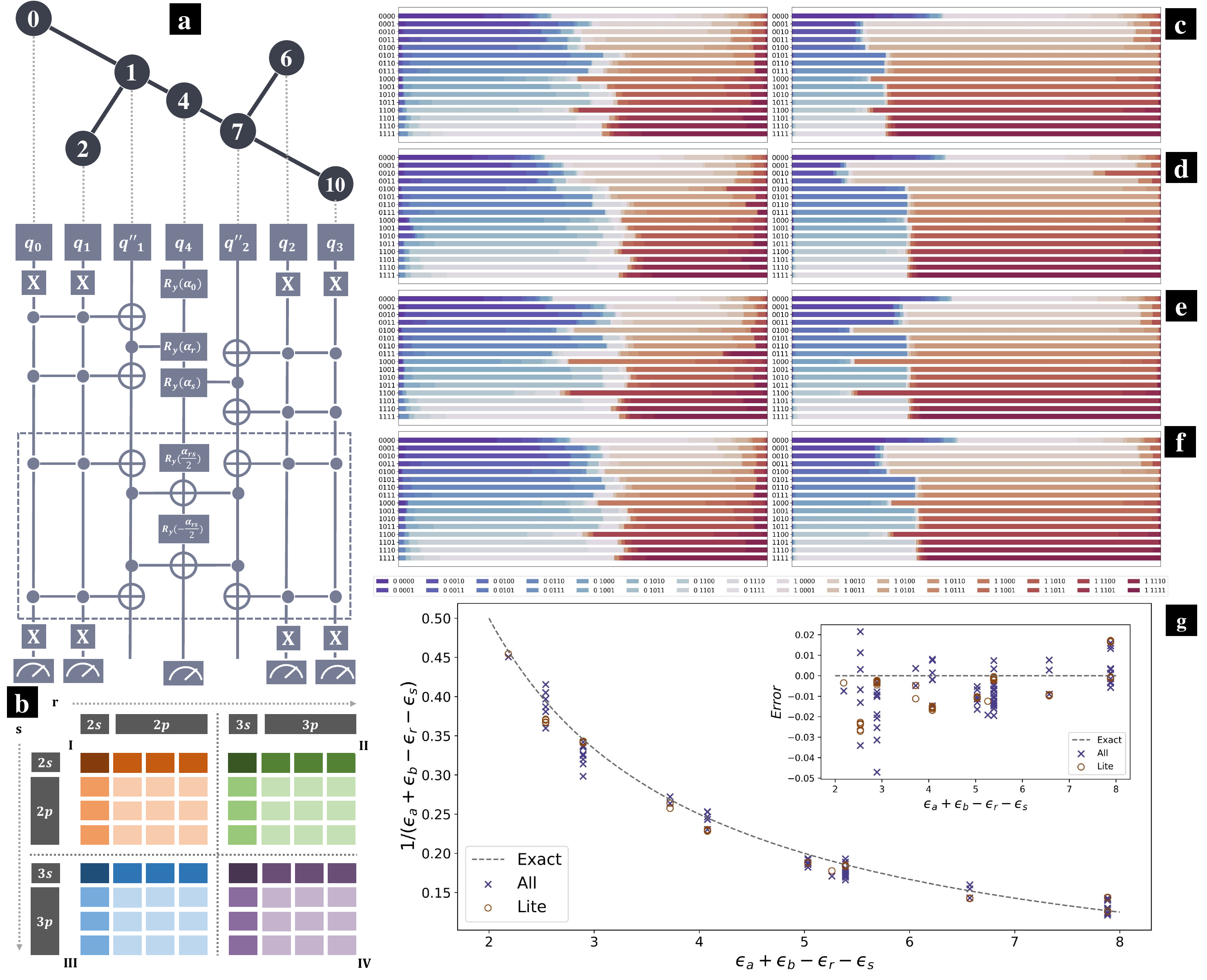}
    \caption{
    {\bf Structure of quantum circuit generating the denominator terms and experiment results from IBM 27-qubit quantum computer.}
    (a) The implementation $U_E$ (lower) and the corresponding qubits on IBM 27-qubit quantum computer (upper).
    (b) A schematic depiction of the matrix expanded by the denominator terms, with indices $r$ and $s$.
    (c,d,e,f) Counts of various outputs for $U_E$ (left column) and $U_E^{lite}$ (right column) with various inputs, corresponding to Part I, II, III, IV of the ERI matrix.
    %For each input, we tested $1e5$ shots.
    (g) The estimated denominators $1/{(\epsilon_a+\epsilon_b-\epsilon_r-\epsilon_s)}$ based on results of $U^{lite}_E$ (brown circles), and the ones based on the corrected results of $U_E$ (blue cross). 
    The dashed line indicate the theoretical predictions.
    Errors are presented in the right upper corner. 
    }
    \label{fig_ue}
\end{figure}

In Fig.(\ref{fig_ue}b) we present a schematic depiction of the matrix expanded by the denominator terms $1/{(\epsilon_a+\epsilon_b-\epsilon_r-\epsilon_s)}$ with indices $r$ and $s$, where boxes with same color indicate the components with same values attributing to the degeneracies in $p$ orbitals.  
Here we divide the matrix into 4 parts, each contains 16 components, raising the requirement of $\log_2 16 = 4$ qubits to represent the relative $r$ and $s$ indices.
The schematic structure of $U_E$ is demonstrated in Fig.(\ref{fig_ue}a), where in the upper part we present the corresponding qubits on IBM 27-qubit quantum computer, `ibm\_aucland'.
$q_{0},q_{1},q_{2}, q_{3}$ are the four qubits representing the orbitals, $q_{0},q_{1}$ for index $r$ and $q_{2},q_{3}$ for index $s$.
In addition, $q_4$ is included for readout, and $q''_1$, $q''_2$ are ancilla qubits in the implementation of Toffoli gates.
Qubits $q_{4}$, $q''_1$, $q''_2$ are initialized at the ground state $|0\rangle$ at beginning, whereas the initial state of $q_{0},q_{1},q_{2}, q_{3}$ corresponds to the relative denominator term to be estimated.
Superpositions in $q_{0},q_{1},q_{2}, q_{3}$ can lead to same superpositions in the outcome.

As presented in Fig.(\ref{fig_ue}a), a single $R_y$ gate is applied on $q_4$ at beginning, generating the $p-p$ components.
Next, two $C^2R_y$ gates are added, preparing the $s-p$ or $p-s$ components.
Finally, a $C^4R_y$ gate is introduced to prepare the $s-s$ component.
Decomposition of the $C^4R_y$ gate is as presented in the dashed box in Fig.(\ref{fig_ue}a).
The Pauli-X gates (or NOT gate, depicted with symbol X) on $q_{0},q_{1},q_{2}, q_{3}$ are included as we are intending to use 0-control instead of 1-control operations.

In the decomposition of a Toffoli gate, there are not only CNOT gates connecting the two control qubits and the target qubit, but also CNOT gate between the two control qubits\cite{nielsen2010quantum}.
Even though, the control qubits are not directly connected on hardware as depicted in Fig.(\ref{fig_ue}a).
Intuitively, we can construct CNOT gates between the two control qubits with assistance of SWAP gates, which however, often leads to extra errors.
Nevertheless, these Toffoli gates can be simplified in the implementation of $U_E$.
Notice that there are no `single' Toffoli gates as shown in Fig.(\ref{fig_ue}a).
Instead, Toffoli gates always appear in pair, without any other operations involving the control qubits between them.
As the inverse of a Toffoli gate is still itself, CNOT gates connecting the two control qubits cancel out in these pairs.
In the decomposition of the Toffoli gates pairs, CNOT gates between the two control qubits are no more necessary (See the supplementary materials or our recent work\cite{li2023toward} for more details).
Therefore, when implementing $U_E$ on real machines, there are only single qubit gates and CNOT gates applied on the physically connected neighbor qubits, eliminating the requirement for extra SWAP gates.
%Eliminating the extra SWAP gates is necessary for accurate estimations, as the extra SWAP gates often lead to deeper circuit with consequent 
%Degeneracies in $p$ orbitals considerably simplified the structure of $U_E$.

We tested $U_E$ with all possible inputs ($|0000\rangle$, $|0001\rangle$, $\cdots$, $|1111\rangle$) on IBM 27-qubit quantum computer `ibm\_aucland', and the results of the four parts are presented in the left columns in Fig.(\ref{fig_ue}c,d,e,f).
For each input, we tested $1\times 10^5$ shots.
We notice that not all operations of $U_E$ are necessary for the certain inputs.
For instance, the intricate $C^4R_y$ gate is designed to work only when the input is $|0000\rangle$, yet the idling operations still raise errors.
To correct the error caused by idling operations, we tested the `lite' version of $U_E$, which only contains the necessary operations, and the results of $U_E^{lite}$ are presented in the right columns in Fig.(\ref{fig_ue}c,d,e,f).
Similarly we tested $1\times 10^5$ shots for each input.
%Errors caused by the idling operations can be estimated by comparing the results of $U_E$ and $U_E^{lite}$.
Corrected estimation is available with Eq.(\ref{eqm_denom}), which roughly cut off the idling operations errors, especially the errors caused by the idling $C^4R_y$ gate.
%, in the results of $U_E$.
In Fig.(\ref{fig_ue}g) we present the estimated denominators $1/{(\epsilon_a+\epsilon_b-\epsilon_r-\epsilon_s)}$ based on results of $U^{lite}_E$ (brown circles), and the ones based on the corrected results of $U_E$ (blue cross).
Errors are presented in the right upper corner of Fig.(\ref{fig_ue}g). 
Both of these estimations are close to the ideal values (dashed line), with maximum error less than $0.05$ (Unit: Hartree${}^{-1}$). 
%We can notice that the corrected results of $U_E$ is more dispersed, which attributes to the randomness of idling operations errors.

%%% FIG UP %%%
\begin{figure}[htbp!]
    \centering
    \includegraphics[width=1.0\textwidth]{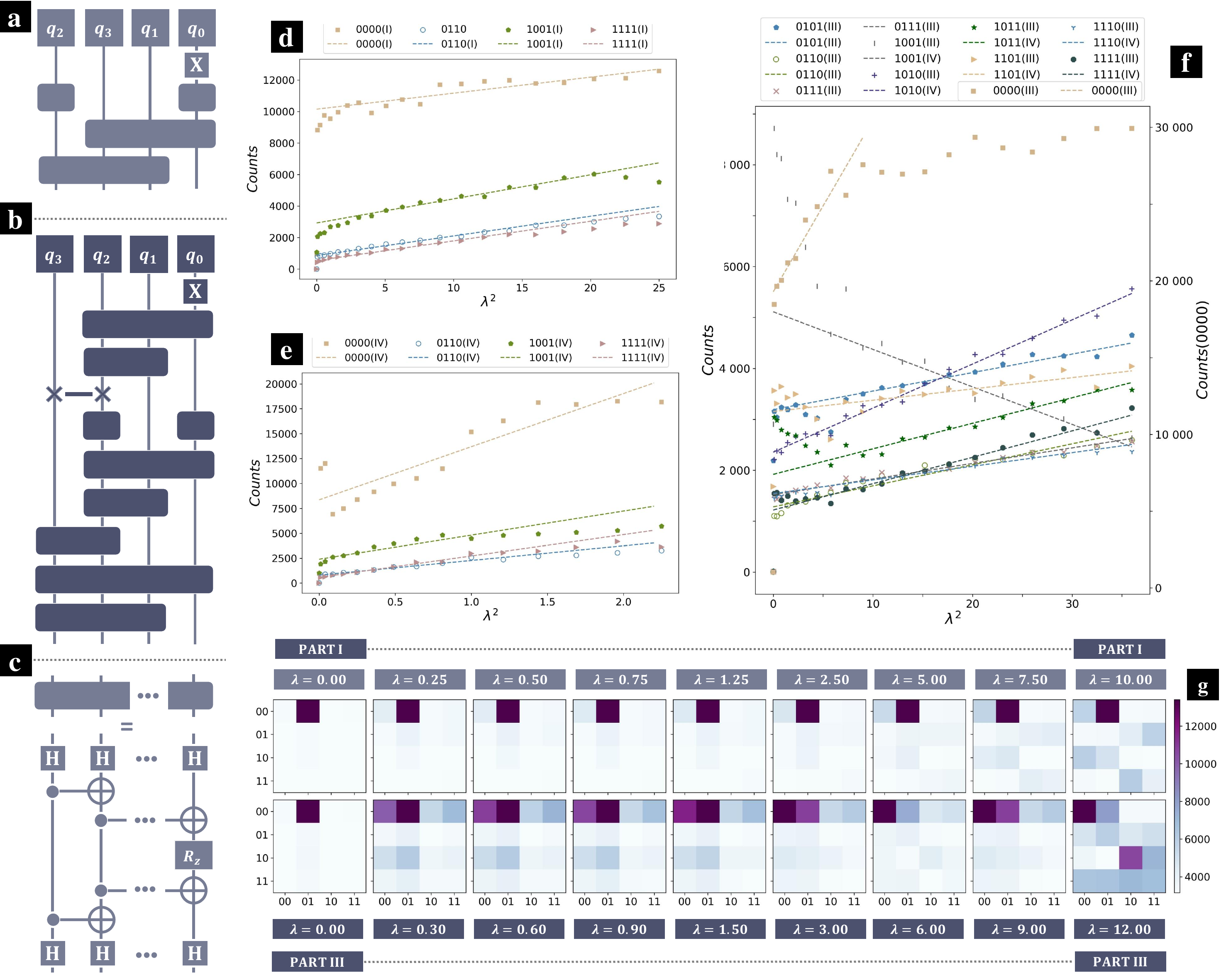}
    \caption{
    {\bf Schematic structure of quantum circuit generating the ERI terms and experiment results from IBM 27-qubit quantum computer.}
    (a) The implementation of $U_{INT}$ for Part I and Part IV.
    (b) The implementation of $U_{INT}$ for Part II and Part III.
    In (a) and (b) the solid boxes indicate operation $\exp(i\lambda\gamma\sigma_x\otimes\cdots\otimes\sigma_x)$ over the covered qubits, where $|\lambda\gamma|\ll1$.
    Decomposition of this operation is depicted in (c).
    The input of $U_{INT}$ is always the ground state $|0000\rangle$.
    In (d,e,f) we present the outcome of $U_{INT}$ with various $\lambda$ values for Part I, Part IV and Part III, where the dots are experiment results collected from IBM 27-qubit quantum computer `ibm\_aucland' (4 qubits involved), and the dashed line are obtained from linear regression with least squares method.
    (g) Detailed results of $U_{INT}$ with various $\lambda$ values for Part I (upper) and Part III (lower).
    }
    \label{fig_up}
\end{figure}

In addition to $U_E$, we also tested $U_{INT}$ on IBM 27-qubit quantum computer, where only 4 qubits are involved.
The mathematical description of $U_{INT}$ can be found in Eq.(\ref{eqm_uint}).
$U_{INT}$ is designed to approximate the ERI components.
In Fig.(\ref{fig_up}a) we present the implementation of $U_{INT}$ for Part I and Part IV, whereas the implementation of $U_{INT}$ for Part II and Part III can be found in (\ref{fig_up}b).
In (a) and (b) the solid boxes indicate operation $\exp(i\lambda\gamma\sigma_x\otimes\cdots\otimes\sigma_x)$ over the covered qubits, where $\gamma$ is the corresponding ERI terms.
Decomposition of this operation is depicted in (\ref{fig_up}c), which is typical in fermionic simulation and Trotterization\cite{ortiz2001quantum, whitfield2011simulation}.
When $|\lambda\gamma|\ll1$, the probability to get result $|n\rangle$ is approximated by $\lambda^2\gamma_n^2$, where the binary form of $n$ is a four-digit number corresponding to the ERI component.
In (\ref{fig_up}d,e,f) we present the outcome of $U_{INT}$ with various $\lambda$ values for Part I, Part IV and Part III, where the dots are experiment results collected from IBM 27-qubit quantum computer `ibm\_aucland' (4 qubits involved).
The dashed lines are obtained from linear regression with least squares method for $\lambda^2$ and counts of various readout, where the slope is proportional to the corresponding ERI component.
There are only 4 non-trivial components in Part I (Part IV) of the ERI matrix, corresponding to output $|0000\rangle$, $|0110\rangle$, $|1001\rangle$ and $|1111\rangle$.
The counts to find these states under a range of $\lambda^2$ are depicted in Fig.(\ref{fig_up}d), whereas the results of Part IV are depicted in Fig.(\ref{fig_up}e).
In Part III (or Part II), there are 10 non-trivial components, and the experiment results are presented in Fig.(\ref{fig_up}f).
For each $U_{INT}$, we tested $1\times 10^5$ shots in total.

In Fig.(\ref{fig_up}d,e,f), we notice that the y-intercept can be far away from 0, which attributes to the noise of $U_{INT}$ with $\lambda=0$ caused by the Toffoli gates pairs as depicted in Fig.(\ref{fig_up}c). 
Meanwhile, for small $\lambda^2$, the scattered points often deviate from the fitted line, as $U_{int}$ with smaller $\lambda$ values are more sensitive to the noises and errors.
On the contrary, the higher order terms in $\exp(i\lambda V)$ is no more insignificant with large $\lambda$ values.
%, as shown in the numerical simulations in Fig.(\ref{fig_main}h).
We can notice the flat plateau around $\lambda^2>10$ for $|0000\rangle$ in Fig.(\ref{fig_up}f), which attributes to the corresponding ERI component that is much greater than the others in Part III.
Detailed output states for Part I (upper), Part III (lower) are depicted in Fig.(\ref{fig_up}g).
For large $\lambda$ values, see $\lambda=10.00$ for Part I, or $\lambda=12.00$ for Part III, unexpected patterns appear in the output, yielding the breakdown of linear approximation.
Briefly, the ERI components can be estimated with the slope of fitted line within appropriate ranges.
We must be careful to avoid both the outliers around small $\lambda$ values and the flat plateau around the large $\lambda$ values.

%%% FIG E2 %%%
\begin{figure}[ht]
    \centering
    \includegraphics[width=0.95\textwidth]{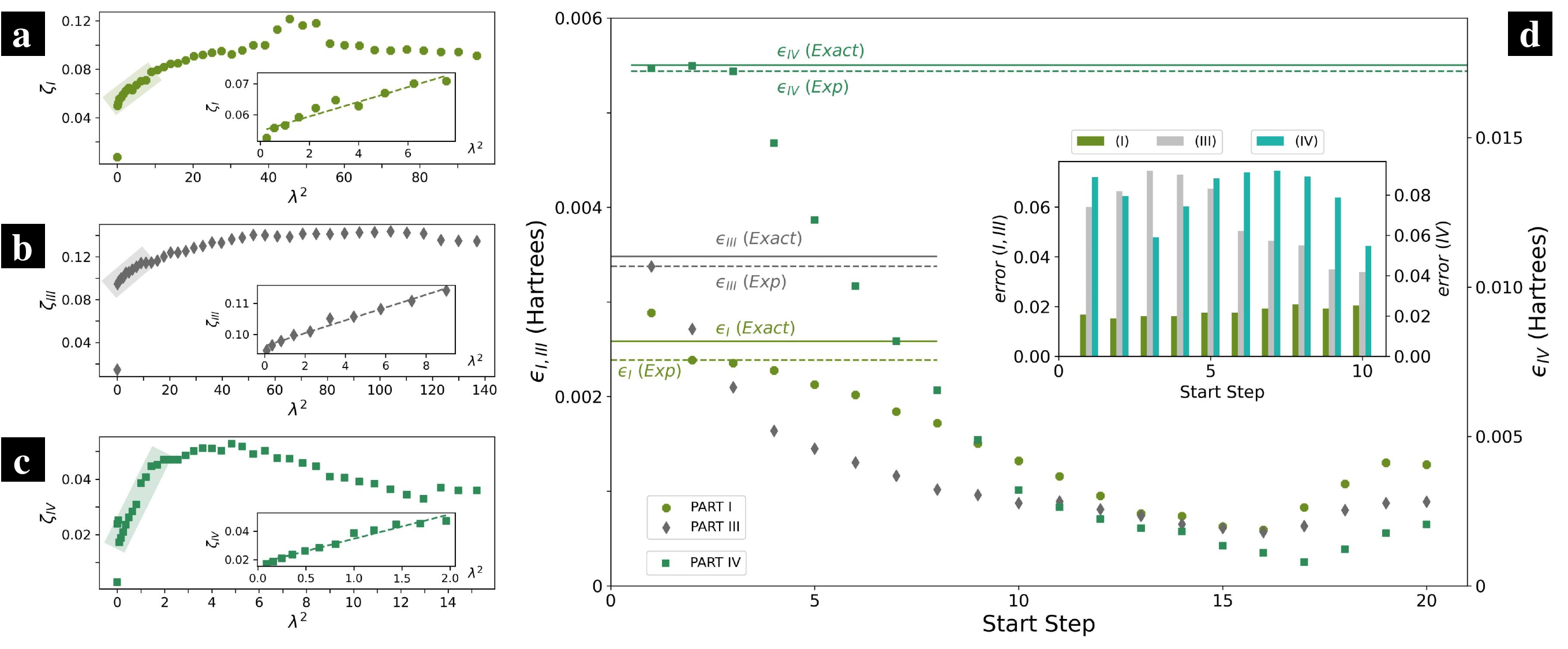}
    \caption{
    {\bf MP2 correlation energy estimated by the proposed quantum circuit.}
    (a,b,c) $\zeta$ values for Part I, Part III and Part IV.
    In the relative subfigures, we present the linear regression with least square method, where the dashed line represents the fitted line.
    (d) MP2 correlation energy against various start step of $\lambda$.
    The step lengths are 0.25, 0.3 and 0.1 for Part I, III and IV.
    The total steps are 10, 10, 12 for Part I, III and IV.
    The least square errors are presented in the right upper corner of (b).
    The solid lines indicate the exact results of $\epsilon_{I,III,IV}$ (Exact), whereas the dashed lines indicate the experiment estimations on quantum devices (Exp).
    }
    \label{fig_e2}
\end{figure}

Experiment results of $U_E$ and $U_{INT}$ enable us to calculate the MP2 correlation energy $E_0^{(2)}$.
In Fig.(\ref{fig_e2}a,b,c) we present the $\zeta$ values for Part I, Part III and Part IV.
Similar to the numerical simulations, $\zeta$ increases linearly against $\lambda^2$ for small $\lambda$ values.
Recalling the simulation results as shown in Fig.(\ref{fig_main}h), it is predicted that there will be a flat plateau after the linear increasing trend.
We can notice that the plateaus shown in Fig.(\ref{fig_e2}a,c) are not the same as the ones in simulation, which is mainly on account of the noise caused by the Toffoli gates in $U_{INT}$.
Even though, this difference does not matter in the estimation of $E_0^{(2)}$.
According to Eq.(\ref{eq_zeta}), $E_0^{(2)}$ can be obtained by calculating the slopes of the fitted lines for the linear increasing part.
To find out the optimal range, we calculated $E_0^{(2)}$ with various start steps, meanwhile the step length and total steps are chosen constant.
The product between step length and total steps is a little less than the length of linear increasing range, so that we can exclude the outliers around small $\lambda$ by choosing the optimal start step that leads to minimum errors in linear regression.
The calculated $E_0^{(2)}$ against start steps are depicted in Fig.(\ref{fig_e2}), and the least square errors (LSE) in linear regression are presented in the right upper corner.
The step lengths are 0.25, 0.3 and 0.1 for Part I, III and IV, whereas the total steps are 10, 10, 12.
The 0th step corresponds to $\lambda=0$.
For Part IV, LSE decreases since the beginning, and reaches local minimum when the 3rd one is chosen as start step.
Therefore, the Part IV contributed correlation energy $\epsilon_{IV}$ is estimated by calculating the slope of fitted line over the highlighted range in Fig.(\ref{fig_e2}c), where the linear regression starts from the 3rd step.
The dash horizontal line colored with seagreen in Fig.(\ref{fig_e2}d) indicates the experiment estimation of $\epsilon_{IV}$ on the proposed quantum circuit, whereas the solid line with same color is the exact result obtained from theoretical calculations.
In Fig.(\ref{fig_e2}a,b), the optimal range are highlighted with light colors.
In the relative subfigures, we present the linear regression with least square method, and the dashed line indicates the fitted line.
in Fig.(\ref{fig_e2}d), the solid lines colored in gray and olivergreen indicate the exact results of $\epsilon_{I,III}$ (Exact), whereas the dashed lines indicate the experiment estimations on quantum devices (Exp).
Recalling Eq.(\ref{eq_zeta}), the absolute value of $E^{(2)}_0$ is thus estimated.
Here we focus on the ground state, ensuring that the denominators $1/{(\epsilon_a+\epsilon_b-\epsilon_r-\epsilon_s)}$ are always negative.
Meanwhile, the numerator terms are always non-negative as shown in Eq.(\ref{eq_mp2}).
Therefore, the MP2 correlation energy is estimated as $E^{(2)}_0 = -\epsilon_I-2\epsilon_{III}-\epsilon_{IV}$.

\begin{table}[H]
    \begin{center}
\begin{tabular}{ |c|c|c|c|c| } 
 \hline
 Unite: Hartrees &$\epsilon_I$ &$\epsilon_{III}$ &$\epsilon_{IV}$ &$E_0^{(2)}$
 \\
 \hline
 \hline
 Exact &0.0025817 &0.0034791 &0.017423 &-0.026963
 \\
 \hline
 Experiment &0.0023838 &0.0033730 &0.017211 &-0.026341
 \\
 \hline
\end{tabular}
\caption{\label{tb_result}Exact results by theoretical calculations (Exact) and experiment estimations on the proposed quantum circuit (Experiment)}
\end{center}
\end{table}

In Tab.(\ref{tb_result}) we list the exact results by theoretical calculations, along with the experiment estimations obtained from quantum devices for $\epsilon_{I}$, $\epsilon_{III}$, $\epsilon_{IV}$ and $E^{(2)}_0$.
$E^{(2)}_0$ estimated by the quantum circuit is very close to the theoretical result, with error rate around $2.3\%$.
As for the four parts, maximum error rate appears when estimating $\epsilon_I$, where the error rate is $7.7\%$.
The theoretical results are calculated by PSI4\cite{parrish2017psi4}, and is exactly the same to the standard one from Computational Chemistry Comparison and Benchmark DataBase (CCCBDB)\cite{johnson2006nist}.

%%% DISCUSSION %%%
In this article, we concentrate on the MPPT correlation energy of a single Helium atom.
In fact, the proposed circuit can be applied to solve more intricate electronic structure problems.
As presented in Fig.(\ref{fig_main}a), the MPPT calculations are implemented in four parts, and the ERI tensor and the energy denominator tensor are both divided into $4\times4$ matrices.
An intricate electronic structure problem often involves more components in the ERI tensor and energy denominator tensor.
By cutting these large-scale tensors into smaller matrices, the complicated problem is simplified into several solvable ones.
$U_{INT}$ is designed to prepare the ERI terms, whereas $U_{E}$ generate the energy denominators.
In Fig.(\ref{fig_main}c) we present the necessary connectivity in implementation of $U_{INT}$, where a solid square represents a single qubit, and the connected squares indicate qubits with physical connection on real devices.
Detailed circuit structure of $U_{INT}$ is depicted in Fig.(\ref{fig_up}a,b).
Meanwhile, the necessary connectivity in implementation of $U_E$ is presented in Fig.(\ref{fig_main}d), corresponding to the detailed circuit depicted in Fig.(\ref{fig_ue}a).
Therefore, it is feasible to accomplish MPPT2 calculations separately on quantum devices with at least 7 qubits as shown in Fig.(\ref{fig_main}e), where the tensors are divided into $4\times4$ matrices.
Furthermore, advanced hardware enables subtle designs.
In Fig.(\ref{fig_main}f) we present the necessary connectivity to implement $U_E$ and $U_{INT}$ simultaneously, corresponding to the schematic circuit design shown in In Fig.(\ref{fig_main}b).
The hollowed squares represent extra qubits involved to build connections among the four qubits implementing $U_{INT}$ as shown in Fig.(\ref{fig_main}d).
The design as shown in Fig.(\ref{fig_main}f) is feasible on quantum computing systems such as Scymore\cite{arute2019quantum} and Zuchongzhi\cite{zhu2022quantum}, where the qubits are assigned in a two-dimensional array.
On the other side, IBM 27-qubit computer enables parallel computing.
As depicted in Fig.(\ref{fig_main}g), three $U_E$ can run in parallel.
Each $U_E$ involves 7 qubits, corresponding to the solid squares colored in navy, purple or blue.

Depth of the proposed circuit is determined by the number of orbitals (MO) involved in the MPPT calculations.
Denote $N$ as the total number of orbitals (MO) in the MPPT calculations, $U_E$ can be decomposed into no more than $\mathcal{O}(N\log^2_2N)$ basic gates (single qubit gates and CNOT gates), and $U_{INT}$ can be decomposed into $\mathcal{O}(N\log_2N)$ basic gates (See the supplementary materials for more details). 
On the contrary, in the traditional MPPT calculations it takes $\mathcal{O}(N^2)$ steps to get the summation as shown in Eq.(\ref{eq_mp2}) (Generally the number of occupied orbitals is much less than the virtual ones).
Therefore, when dealing with intricate electronic structure problems with $N\gg1$, the proposed circuit could lead to considerable speedup comparing to the traditional MPPT methods, yet which is still in need of substantial advancements in hardware developments.

\section*{Conclusions}
%%% CONCLUSION %%%
As one of the most promising applications of quantum computing, it has been attracting enormous enthusiasms to solve electronic structure problems on quantum devices.
Illustrated by the recent successful quantum simulations within the Hartree-Fock approximation, we proposed a general quantum circuit for MPPT calculations, which is a powerful post-Hartree-Fock method based on Rayleigh–Schrödinger perturbation theory.

There are two major tasks to calculate the MP2 correlation energy on quantum devices, estimating the denominators determined by the unperturbed energy levels, and estimating the nominators determined by the ERI tensor, as shown in Eq.(\ref{eq_mp2}).
To address these two issues, we designed $U_E$ preparing the denominators, and $U_{INT}$ approximating the ERIs.
Due to the limitation of NISQ devices, these tensors are cut into smaller matrices, and the MPPT calculation is simplified into several solvable problems.
Nevertheless, we then implemented the proposed circuit on IBM 27-qubit quantum computer, and estimated the MP2 correlation energy for the ground state of Helium atom($\ce{{}^4He}$), where the total error rate is around $2.3\%$.

In summary, a general quantum circuit to calculate the MP2 correlation energy with the given Hartree-Fock results is proposed in this article.
The proposed circuit is feasible on NISQ devices, and is capable of estimating the MP2 correlation energy accurately.
Our work provides an approach to implement MPPT calculations on quantum devices, enriching the toolkit solving electronic structure problems
and smoothing the path to implement other {\it{ab initio}} methods, especially the ones based on perturbation theory, on quantum hardware.
%We implemented the proposed circuit on IBM 27-qubit quantum computer, and estimated the MP2 correlation energy for the ground state of Helium atom($\ce{{}^4He}$), where the total error rate is around $2.3\%$.
%Experiment results indicate that the proposed quantum circuit is a feasible and promising approach to solve the electronic structure problems on near term quantum devices.
%Our work smooths the path to implement other post Hartree-Fock methods, or even {\it{ab initio}} quantum chemistry toolkit on quantum hardware.
%Our work includes the quantum-enhanced MPPT calculations into the toolkit of computational chemistry, 

%%% Methods %%%
\section*{Methods}
\subsection*{Preliminaries to Møller–Plesset perturbation theory}
%The second order MP (MP2) correlation energy for ground state is\cite{szabo2012modern}
%where $a,b$ indicates the occupied orbitals, $r,s$ indicates virtual orbitals, and $\epsilon$ is the orbital energies obtained from Hartree–Fock calculations.
The antisymmetrized two electron integral is defined as
\begin{equation}
    \langle ab||rs\rangle = \langle ab|rs\rangle - \langle ab|sr\rangle
    \label{eqm_anti}
\end{equation}
and the electron repulsion integral (ERI) under physicists notation is 
\begin{equation}
    \langle ab|rs\rangle =  \int \phi^*_a(x_1)\phi^*_b(x_2)
    \frac{1}{r_{12}}\phi_r(x_1)\phi_s(x_2)dx_1dx_2
    \label{eqm_eri_mo}
\end{equation}
where $\phi_a(x)$ is the spatial orbital.
%Here we study the MP2 correlation enenrgy for Helium, where the two occupied obitals are both $1s$ orbitals, ensuring that
%\begin{equation}
%    E_0^{(2)} = \sum_{abrs}^{N/2}
%    \frac{\langle ab|rs\rangle\langle rs|ab\rangle}
%    {\epsilon_a+\epsilon_b-\epsilon_r-\epsilon_s}
%\end{equation}

\subsection*{Decomposition of the key operations}
$U_E$ is defined as,
\begin{equation}
    U_E=\prod_{x=0}^{N_{tot}-1}
    \left\{
    \left[
    \bigotimes_{j=1}^{Q_{tot}}
    \left(
    (1-x_j)|0\rangle\langle 0|+|1\rangle\langle 1|
    \right)
    \right]
    \otimes
    R_y(\alpha^{sqrt}_x)
    +
    %\left[
    %\bigotimes_{j=1}^{Q_{tot}}
    %\left(
    %|0\rangle\langle 0|+(1-x_j)|1\rangle\langle 1|
    %\right)
    %\right]
    \left[
    I^{\otimes Q_{tot}}-
    \bigotimes_{j=1}^{Q_{tot}}
    \left(
    (1-x_j)|0\rangle\langle 0|+|1\rangle\langle 1|
    \right)
    \right]
    \otimes
    I
    \right\}
    \label{eqm_ue}
\end{equation}
where $N_{tot}$ is the total number of orbitals, $Q_{tot}$ is the total number of qubits, $x$ is an integer indicating the MO, corresponding to the indices $a,b,r,s$, and $x_j\in\{0,1\}$ is the $j-th$ digit in the binary form of $x$.
In the description as shown in Eq.(\ref{eqm_ue}), the $C^nR_y$ gates are 1-control gates.
If 0-control gates are employed instead, we can get the mathematical definition just by exchanging the $R_y$ gates and identity gates $I$ to Eq.(\ref{eqm_ue}).
$\alpha^{sqrt}$ is chosen to guarantee that
\begin{equation}
    \sum_{y\in Y(x)}\alpha^{sqrt}_y=
    \arccos{\left(1-
    \frac{2C_e}{\epsilon_a+\epsilon_b-\epsilon_r-\epsilon_s}
    \right)}
\end{equation}
where $C_e$ is a constant ensuring that $C_e/(\epsilon_a+\epsilon_b-\epsilon_r-\epsilon_s)\in[0, 1]$, and the set $Y(x)$ is defined as
\begin{equation}
    Y(x)=
    \left\{y
    \mid
    y=\sum_{j=0}^{N_{tot}-1}2^jy_j, y_j\in\{x_j,0\}
    \right\}
    \label{eqm_set_y}
\end{equation}
$y_j$ is the $j-th$ digit in the binary form of $y$.

$U_E$ ensures that
\begin{equation}
    U_E(|0\rangle_{q'}\otimes|abrs\rangle_{q})
    =
    \left(
    \sqrt{1-\left|\frac{C_e^{sqrt}}{\epsilon_a+\epsilon_b-\epsilon_r-\epsilon_s}\right|}
    |0\rangle_{q'}
    +
    \sqrt{\frac{C_e^{sqrt}}{\epsilon_a+\epsilon_b-\epsilon_r-\epsilon_s}}
    |1\rangle_{q'}
    \right)
    \otimes|abrs\rangle_{q}
    \label{eqm_ue_sqrt}
\end{equation}

The definition of $U_{INT}$ in first order Trotter decomposition is
\begin{equation}
    U_{INT}(\lambda)=
    \prod_{x=0}^{N_{tot}-1}
    \left\{
    \exp{
    \left(
    i\lambda
    \gamma_{x}
    \bigotimes_{j=0}^{Q_{tot}}
    \sigma_{x_j\oplus y_j}
    \right)
    }
    \right\}
    \cdot
    \bigotimes_{j=0}^{Q_{tot}}
    \sigma_{y_j}
    \label{eqm_uint}
\end{equation}
where $\sigma_1$ is the Pauli-X gate, $\sigma_0$ is identity gate, $\gamma_y=0$, and we denote the ERI in MOs as 
\begin{equation}
    \gamma_{abrs}=
    \langle ab|rs\rangle
\end{equation}

For $|\lambda\gamma_x|\ll 1$, $U_{INT}$ ensures that
\begin{equation}
    U_{INT}(\lambda)|0\rangle
    =
    |y\rangle
    +i\lambda\sum_{x\neq y}^{N_{tot}}\gamma_x|x\rangle
    +\mathcal{O}(\lambda^2)
    \label{eqm_uint_approx}
\end{equation}

\subsection*{Estimation of the denominators}
The denominator terms are estimated as
\begin{equation}
    \left|\frac{C_e}{\epsilon_a+\epsilon_b-\epsilon_r-\epsilon_s}\right| 
    =
    \left(\frac{1}{15}\sum_{n=1}^{16}
    \frac{\tilde{P}_n^n+p^n_{n+16}}{\tilde{P}_n^n + \tilde{P}_{n+16}^n}
    \right)^{-1}\cdot
    \frac{p^x_{x+16}}{p_x^x + p^x_{x+16}}
    \label{eqm_denom}
\end{equation}
where $x$ is an integer corresponding to the state of $q'$ and $q$, $p_{x}^y$ is the count to find output $x$ with input $y$ with operation $U_E^{all}$, $\tilde{P}$ corresponds to the count under operation $U_E^{lite}$.
Here we focus on the simple Helium atom, and 4 $q$ qubits are involved in $U_{INT}$.
Thus, state $|x\rangle$ represents $|0\rangle_{q'}\otimes|(x)_{bin}\rangle_{q}$, whereas state $|x+16\rangle$ represents $|1\rangle_{q'}\otimes|(x)_{bin}\rangle_{q}$, where $(x)_{bin}$ indicates the binary form of integer $x$, and subscripts $q$ or $q'$ indicate the corresponding qubits.

More details of the proposed circuit can be found in the supplementary materials.

\section*{Data availability}
Source data are available for this paper.
All other data that support the plots within this paper and other findings of this study are available from the corresponding author upon reasonable request.

\section*{Acknowledgements}
We would like to thank Dr. Barbara Jones for many useful discussions on using perturbation calculations on quantum devices, and Dr. Zhang-Run Xu, Dr. Yue Wang for helpful discussions on MPPT calculations.
S.K. acknowledge the financial support of the National Science Foundation under Award 2124511, CCI Phase I: NSF Center for Quantum Dynamics on Modular Quantum Devices (CQDMQD).
M.S.  and S.K.  acknowledge the use of IBM Quantum services for this work. The views expressed are those of the authors and do not reflect the official policy or position of IBM or the IBM Quantum team.

\section*{Competing interests}
The authors declare no competing interests.

\section*{Author contributions}
J.L. and S.K. conceived and designed the study.
J.L. developed the quantum circuits.
J.L. and X.G. prepared the QASM codes for quantum computing.
M.S. implemented the quantum circuits on IBM quantum computing platforms. 
All authors discussed the results and wrote the paper. 
%\subsection*{Corresponding authors}
%Correspondence to Junxu Li and Sabre Kais.

\bibliography{ref}

%%%%%%%%%%%%%%%%%%%%%%%%%%%%%%%%%%%%%%%%%%%%%%%%%%%
%%%%       SUPPLEMENTARY MATERIALS              %%%
%%%%%%%%%%%%%%%%%%%%%%%%%%%%%%%%%%%%%%%%%%%%%%%%%%%
\newpage
\section*{Supplementary Materials}
\newcommand{\beginsupplement}{
    \setcounter{section}{0}
    \renewcommand{\thesection}{S\arabic{section}}
    \setcounter{equation}{0}
    \renewcommand{\theequation}{S\arabic{equation}}
    \setcounter{figure}{0}
    \renewcommand{\thefigure}{S\arabic{figure}}}
\beginsupplement
\section{Preliminaries to Møller–Plesset perturbation theory}
\label{Preliminaries}

Reliable {\it ab initio} electronic structure calculations require an accurate treatment of the many-particle effects.
In 1934, Møller and Plesset proposed a perturbation treatment of atoms and
molecules\cite{moller1934note}, which improves on the Hartree–Fock method\cite{hartree1928wave, fock1930naherungsmethode} by means of Rayleigh–Schrödinger perturbation theory (RS-PT)\cite{schrodinger1926quantisierung}, where the perturbation $H'$ is the difference between the true molecular electronic Hamiltonian $H$ and the Hartree–Fock Hamiltonian $H_{HF}$\cite{levine2009quantum},
\begin{equation}
    \hat{H'}=\hat{H}-\hat{H}_{HF}
    =\sum_{l}\sum_{m>l}\frac{1}{r_{ml}} 
    - \sum_{m=1}^n\sum_{j=1}^n\left[
    \hat{J}_j(m)-\hat{K}_j(m)
    \right]
\end{equation}
The Coulomb operator $\hat{J}_j(m)$ and the exchange operator $\hat{K}_j(m)$ are defined by
\begin{equation}
    \hat{J}_j(1)f(1) = f(1)\int|\varphi_j(2)|^2\frac{1}{r_{12}}dx_2
\end{equation}
\begin{equation}
    \hat{K}_j(1)f(1) = \varphi_j(1)\int\varphi^*_j(2)f(2)\frac{1}{r_{12}}dx_2
\end{equation}
where $\varphi_j(k)$ is a branch of one-electron wave functions, often called the Hartree–Fock molecular orbitals.
This treatment is later called Møller–Plesset (MP) perturbation theory.

The first order MP correlation energy is 0, and the second order MP (MP2) correlation energy for ground state is\cite{szabo2012modern}
\begin{equation}
    E_0^{(2)} = \sum_{a<b}\sum_{r<s} 
    \frac{|\langle ab||rs\rangle|^2}{\epsilon_a+\epsilon_b-\epsilon_r-\epsilon_s}
    \label{eqs_mp2}
\end{equation}
where $a,b$ indicate the occupied orbitals, $r,s$ indicate the virtual orbitals, and $\epsilon$ is the orbital energies from Hartree–Fock calculations.
The antisymmetrized two electron integral is defined as
\begin{equation}
    \langle ab||rs\rangle = \langle ab|rs\rangle - \langle ab|sr\rangle
    \label{eq_anti}
\end{equation}
and the physicists notation is 
\begin{equation}
    \langle ab|rs\rangle =  \int \phi^*_a(x_1)\phi^*_b(x_2)
    \frac{1}{r_{12}}\phi_r(x_1)\phi_s(x_2)dx_1dx_2
    \label{eq_eri_mo}
\end{equation}
$\phi_a(x)$ is the spatial orbitals that can be obtained from Hartree-Fock calculations,
\begin{equation}
    \phi_a(x) = \sum_k c_{ka} \chi_k(x)
    \label{expansion_phi}
\end{equation}
where $\chi_k(x)$ is the chosen basis.
%In computational chemistry, $\chi$ is often denoted as atomic orbitals (AO), whereas $\phi$ is denoted as molecular orbitals (MO).
For a closed-shell system, the MP2 correlation energy can be written in terms of sum over all spatial orbitals\cite{szabo2012modern},
\begin{equation}
    E_0^{(2)} = 2\sum_{abrs}^{N/2}
    \frac{\langle ab|rs\rangle\langle rs|ab\rangle}
    {\epsilon_a+\epsilon_b-\epsilon_r-\epsilon_s}
    -
    \sum_{abrs}^{N/2}\frac{\langle ab|rs\rangle\langle rs|ba\rangle}
    {\epsilon_a+\epsilon_b-\epsilon_r-\epsilon_s}
\end{equation}
Furthermore, the accuracy of MP2 calculations can be improved by semi-empirically scaling the opposite-spin (OS) and same-spin (SS) correlation components with separate scaling factors, as shown by Grimme\cite{grimme2003improved}.

In the main article we study the MP2 correlation enenrgy for Helium, where the two occupied obitals are both $1s$ orbitals, ensuring that
\begin{equation}
    E_0^{(2)} = \sum_{abrs}^{N/2}
    \frac{\langle ab|rs\rangle\langle rs|ab\rangle}
    {\epsilon_a+\epsilon_b-\epsilon_r-\epsilon_s}
\end{equation}
In the following discussion, we use the basis `aug-cc-pvdz'\cite{kendall1992electron}, including $1s2s2p3s3p$ orbitals.
In classical calculations, the properties such as molecular orbitals (MO), electronic repulsion integrals (ERI) are all obtained from PSI4\cite{parrish2017psi4}.
The Hartree–Fock energy $E_{HF}$ and MP2 correlation energy $E_{MP2}$ of neutral helium atom obtained from PSI4 are (Units: Hartrees)
\begin{equation*}
    \begin{split}
        E_{HF} = -2.8557047
        \\
        E_{MP2} = -2.8826672
    \end{split}
\end{equation*}
and
\begin{equation*}
    E_0^{(2)} = -0.0269625
\end{equation*}
which are exactly the same to the standard ones from Computational Chemistry Comparison and Benchmark DataBase (CCCBDB)\cite{johnson2006nist}.

In the MP2 calculations, totally 9 spatial orbitals $\{\phi_a(x), a=1,2,\cdots, 9\}$ are taken into account, corresponding to $1s,2s,2px,2py,2pz, 3s, 3px,3py,3pz$ orbitals.
The spatial orbital can be written as a linear combination of the chosen `aug-cc-pvdz' basis $\{\chi_k, k=1,2,\cdots, 9\}$, as shown in Eq.(\ref{expansion_phi}).
In computational chemistry, these basis function $\chi$ and atomic orbital (AO) are often used interchangeably, yet the basis functions are usually not the real AOs.
Meanwhile, the spatial orbital $\phi$ are often called molecular orbital(MO), though the one-electron function $\phi$ is not a true molecular orbital.
In the following discussion, AOs refer to the basis function $\chi$, with subscripts $k,l,m,n$, while MOs refer to the spatial orbitals $\phi$, with subscripts $a,b,r,s$.
In this context, $\langle ab|rs\rangle$ represents the ERI of MOs, whereas $\langle kl|mn\rangle$ indicates the ones of AOs.
The ERI in MOs can be transformed from the ones in AOs,
\begin{equation}
    \langle ab|rs\rangle
    =\sum_k\sum_l\sum_m\sum_n
    c^*_{ka}c^*_{lb}c_{mr}c_{ns}
    \langle kl|mn\rangle
\label{eq_expansion_eri}
\end{equation}
The AOs are well studied in the past decades, and some typical AOs, such as the Gaussian Type Orbitals (GTO)\cite{boys1950electronic}, could lead to considerable speedup in the integral calculations.
Therefore, it is often convenient to calculate ERI of MOs with Eq.(\ref{eq_expansion_eri}).

\section{Quantum circuits for MP2 correlation energy calculations}

In the main article we propose to calculate the MP2 correlation energy on a NISQ quantum device\cite{preskill2018quantum}.
To implement the MP2 calculations on a quantum computer, there are three main tasks:

%\noindent\space\space\space\space
{\bf 1.} Operations generating the denominators $1/{(\epsilon_a+\epsilon_b-\epsilon_r-\epsilon_s)}$;

%\noindent\space\space\space\space
{\bf 2.} Operations estimating the ERI tensor $\langle ab|rs\rangle$;

%\noindent\space\space\space\space
%{\bf 3.} Operations transforming the atomic orbitals (AO) into molecular orbitals (MO);

{\bf 3.} Operations estimating the difference.

In the main article we focus on the single Helium atom (\ce{{}^4He}), where task 3 is not necessary.
Herein we present the detailed circuits harnessed for all of the three major tasks above.
Meanwhile, there are also the mathematical descriptions and thorough discussion about their functions.
We will also demonstrate the more general design, along with discussion about the universal cases and analysis of time complexity.
Additionally, we will also present design of operations transforming the atomic orbitals (AO) into molecular orbitals (MO).

There are branches of intricate mathematical notations in the succeeding sections.
For clarification, we list most notations as follows.
Subscripts $a,b,r,s$ always indicate the MOs $\phi$, whereas $k,l,m,n$ indicate the AOs $\chi$.
The numbers of orbitals are denoted as $N$, and the numbers of qubits are denoted as $Q$.
The total number of shots in measurement is denoted as $M$.
The qubits representing orbitals are denoted as $q$, whereas the ancilla ones are denoted as $q'$ and $q''$.
$q'$ qubits are included for readout, and $q''$ are employed to implement multi-controller operations.
Capital notation $C$ refer to a variety of constants, while $c$ always indicate the expansion coefficients in Eq.(\ref{expansion_phi}).

\subsection{Circuits generating the denominator terms}
\label{Circuits generating the denominator terms}

\begin{figure}[ht]
    \centering
    \includegraphics[width=0.95\textwidth]{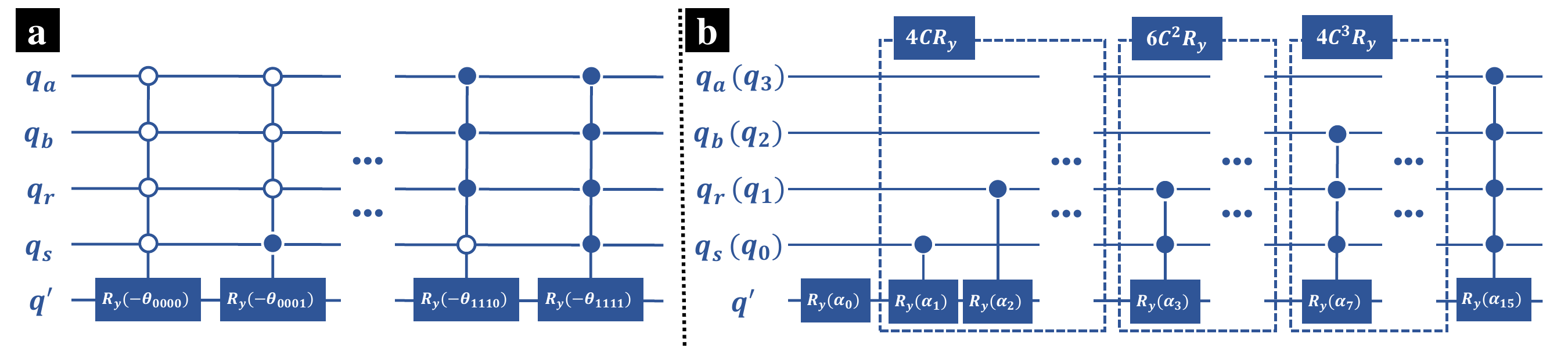}
    \caption{
    {\bf Schematic structure of quantum circuit generating the denominator terms.}
    (a) The intuitive implementation $U'_e$ that only contains multi-controller $R_y$ gates.
    (b) The improved design $U_e$, where for simplicity, there are only four $q$ qubits, corresponding to the $a,b,r,s$ orbitals.
    }
    \label{figs_ue1}
\end{figure}

Denote the quantum circuits as $U_e$ that generate the denominator terms.
The orbital energy $\epsilon$ correspond to the MOs, and are obtained by solving the Roothaan equations by an iterative process\cite{levine2009quantum}.
Thereby, $U_e$ should act on the qubits $q_a, q_b, q_r, q_s$ that represent the MOs.
In addition, $U_e$ also acts on an ancilla qubit $q'$ initialized as ground state $|0\rangle$, with which the denominator terms are prepared as expected.
Mathematically, we have, 
\begin{equation}
    U_e(|abrs\rangle\otimes|0\rangle)
    =
    |abrs\rangle\otimes
    \left(
    \sqrt{1-\left|\frac{C_e}{\epsilon_a+\epsilon_b-\epsilon_r-\epsilon_s}\right|^2}
    |0\rangle
    +
    \frac{C_e}{\epsilon_a+\epsilon_b-\epsilon_r-\epsilon_s}
    |1\rangle
    \right)
    \label{eq_ue_aim}
\end{equation}
where $C_e$ and is a constant ensuring that $|C_e/(\epsilon_a+\epsilon_b-\epsilon_r-\epsilon_s)|\in[0,1]$.
Intuitively, $U_e$ can be implemented with a branch of multi-controller operations, as depicted in Fig.(\ref{figs_ue1}a).
The corresponding mathematical description is
\begin{equation}
    U'_e
    =
    \prod_{abrs}|abrs\rangle\langle abrs|\otimes R_y(\theta_{abrs})
\end{equation}
where
\begin{equation}
    \theta_{abrs} = 2\arcsin{
    \left(
    \frac{C_e}{\epsilon_a+\epsilon_b-\epsilon_r-\epsilon_s}
    \right)
    }
\end{equation}
$C_e$ is the same constant in Eq.(\ref{eq_ue_aim}).
Here we denote this implementation as $U'_e$ to avoid confusion with the improved circuit $U_e$.

Denote the total number of orbitals as $N_{tot}$, we have $N_{tot} = N_a+N_b +N_r+N_s$, where $N_a, N_b$ are number of occupied orbitals, and $N_r, N_s$ of virtual orbitals.
These orbitals are mapped to $Q_{tot}$ $q$ qubits, where $Q_{tot}=Q_a+Q_b+Q_r+Q_s$, and $Q_{a,b,r,s}=\lceil\log_2N_{a,b,r,s}\rceil$ are numbers of qubits representing the corresponding orbitals.
For each orbital, there exists a $C^{Q_{tot}}R_y$ gate, where $C^{Q_{tot}}$ indicates that there are ${Q_{tot}}$ control qubits in the multi-controller gate.
A $C^{Q_{tot}}R_y$ gate can be decomposed into $\mathcal{O}(Q^2_{tot})$ CNOT gates and single qubit gates\cite{barenco1995elementary}.
In other words, it consumes $\mathcal{O}(Q^2_{tot})$ basic operations to build up a $C^{Q_{tot}}R_y$ gate.
Notice that $\mathcal{O}({Q_{tot}})=\mathcal{O}({\log N_{tot}})$, thereby the time complexity of $U'_e$ is $\mathcal{O}(N_{tot}Q^2_{tot})=\mathcal{O}(N_{tot}\log^2N_{tot})$.

In the intuitive design $U'_e$, only the multi-controller gates with $Q_{tot}$ control qubits are harnessed, yet in fact, the multi-controller gates with less control qubits can also take place in the preparation of denominator terms.
An improved implementation of $U_e$ can be written as,
\begin{equation}
    U_e=\prod_{x=0}^{N_{tot}-1}
    \left\{
    \left[
    \bigotimes_{j=1}^{Q_{tot}}
    \left(
    (1-x_j)|0\rangle\langle 0|+|1\rangle\langle 1|
    \right)
    \right]
    \otimes
    R_y(\alpha_x)
    +
    %\left[
    %\bigotimes_{j=1}^{Q_{tot}}
    %\left(
    %|0\rangle\langle 0|+(1-x_j)|1\rangle\langle 1|
    %\right)
    %\right]
    \left[
    I^{\bigotimes Q_{tot}}-
    \bigotimes_{j=1}^{Q_{tot}}
    \left(
    (1-x_j)|0\rangle\langle 0|+|1\rangle\langle 1|
    \right)
    \right]
    \otimes
    I
    \right\}
    \label{eq_ue_improve}
\end{equation}
where $x$ is an integer indicating the MO corresponding to indices $a,b,r,s$, and $x_j\in\{0,1\}$ is the $j-th$ digit in the corresponding binary form of $x$.
Order of the factors in Eq.(\ref{eq_ue_improve}) does not matter, as all of the multi-controller $R_y$ gates commute.
In the general design of $U_e$, single qubit $R_y$ gate is firstly employed to generate the denominator terms as shown in Eq.(\ref{eq_ue_aim}), then the $CR_y$ gates, then the $C^2R_y$ gates and so on, where $q'$ qubit is always the target, and $q$ qubits are the control qubits.
In Fig.(\ref{figs_ue1}b) the schematic structure of $U_e$ is presented, where for simplicity, there are only 4 $q$ qubits, corresponding to the $a,b,r,s$ orbitals.
A simple example might be fruitful to illustrate Eq.(\ref{eq_ue_improve}).
Consider the case of orbital $a=b=s=0, r=1$ corresponding to $x=2$.
The binary form of $x=2$ is $0010$, where there are one digit as $1$, leading to a $CR_y$ gate, and $q_3$ is the control qubit.
For the simplest case where there are 4 $q$ qubits representing the orbitals, there are totally 1 $R_y$ gate, 4 $CR_y$ gates, 6 $C^2R_y$ gates, 4 $C^3R_y$ gates and 1 $C^4R_y$ gate.

Another issue arises when figuring out the values of $\alpha$ in $U_e$.
Differ from the direct one-to-one connection between $\theta$ and the denominator terms, most $\alpha$ contribute to more than one denominator terms (The only exception is $\alpha_{N_{tot}-1}$, which only makes contribution to the highest-order term).
For instance, $\alpha_0$ contributes to all terms.
Constraints to the $\alpha$ values can be written as
\begin{equation}
    \sum_{y\in Y(x)}\alpha_y=
    2\arcsin{\left(
    \frac{C_e}{\epsilon_a+\epsilon_b-\epsilon_r-\epsilon_s}
    \right)}
\end{equation}
where $x$ is the integer representing orbitals $a,b,r,s$, and the set $Y(x)$ is
\begin{equation}
    Y(x)=
    \left\{y
    \mid
    y=\sum_{j=0}^{N_{tot}-1}2^jy_j, y_j\in\{x_j,0\}
    \right\}
    \label{eq_set_y}
\end{equation}
$x_j$, $y_j$ are digits in the binary forms of $x$, $y$.

The design of $U_e$ is intricate but nevertheless fruitful as the employment of various multi-controller gates saves time in quantum circuit implementation.
Theoretically, in the quantum circuit $U_e$ designed for MP2 calculation with $N_{tot}$ orbitals, the number of $C^jR_y$ gates is $\binom{Q_{tot}}{j}$, and $j=0$ refers to the single qubit $R_y$ gate.
Recalling that a $C^{j}R_y$ gate can be decomposed into $\mathcal{O}(j^2)$ CNOT gates and single qubit gates\cite{barenco1995elementary}.
Then the total time complexity of $U_e$ is,
\begin{equation} 
    \mathcal{O}
    \left(
    \sum_{j=0}^{Q_{tot}}
    \binom{Q_{tot}}{j}j^2
    \right)
\end{equation}
which infers that $U_e$ consumes less than $U'_e$.
Therefore, it is more instructive to generate the denominator terms with $U_e$ instead of $U'_e$, especially when there are plenty of MOs involved.

Sometimes we prefer to prepare the denominator terms into the measurement probabilities.
Then an alternative denoted as $U_e^{sqrt}$ might be helpful, which is denoted as $U_E$ in the main article.
\begin{equation}
    U^{sqrt}_e(|abrs\rangle\otimes|0\rangle)
    =
    |abrs\rangle\otimes
    \left(
    \sqrt{1-\left|\frac{C_e^{sqrt}}{\epsilon_a+\epsilon_b-\epsilon_r-\epsilon_s}\right|}
    |0\rangle
    +
    \sqrt{\frac{C_e^{sqrt}}{\epsilon_a+\epsilon_b-\epsilon_r-\epsilon_s}}
    |1\rangle
    \right)
    \label{eq_u_e_sqrt}
\end{equation}
and $C^{sqrt}_e/(\epsilon_a+\epsilon_b-\epsilon_r-\epsilon_s)\in[0,1]$.
$U_e$ directly generate the $1/(\epsilon_a+\epsilon_b-\epsilon_r-\epsilon_s)$ terms, whereas $U^{sqrt}_e$ introduces the $\sqrt{1/(\epsilon_a+\epsilon_b-\epsilon_r-\epsilon_s)}$ terms.
The corresponding constraints are
\begin{equation}
    \theta^{sqrt}_{abrs} = \arccos{
    \left(
    1-
    \frac{2C_e}{\epsilon_a+\epsilon_b-\epsilon_r-\epsilon_s}
    \right)
    }
\end{equation}
and
\begin{equation}
    \sum_{y\in Y(x)}\alpha^{sqrt}_y=
    \arccos{\left(1-
    \frac{2C_e}{\epsilon_a+\epsilon_b-\epsilon_r-\epsilon_s}
    \right)}
\end{equation}
ensuring that
\begin{equation}
    \sin^2\left(\frac{\theta^{sqrt}_{abrs}}{2}\right)
    =
    \frac{C_e}{\epsilon_a+\epsilon_b-\epsilon_r-\epsilon_s}
\end{equation}
and
\begin{equation}
    \sin^2\left(\frac{1}{2}\sum_{y\in Y(x)}\alpha^{sqrt}_y\right)
    =
    \frac{C_e}{\epsilon_a+\epsilon_b-\epsilon_r-\epsilon_s}
\end{equation}
where $Y(y)$ is the same set as in Eq.(\ref{eq_set_y}).
The superscripts $sqrt$ indicates that we are preparing the square root of the denominator terms.
In the discussion above, the $C^nR_y$ gates are 1-control gates.
If 0-control gates are employed instead, we can get the mathematical definition just by exchanging the $R_y$ gates and identity gates $I$.

\subsection{Circuits estimating ERI}
\label{Circuits estimating ERI}

In this subsection we focus on the quantum circuits generating the ERI in MOs.
One approach is to prepare the ERI in AOs with multi controller gates, which is similar to $U_e'$ and $U_e$.
Denote the quantum circuit that estimates the ERI as $U_{int}$, where the subscript $int$ indicates `integrals'.
Recalling Eq.(\ref{eq_eri_mo}), the ERI in MOs, denoted as $\langle ab|rs\rangle$ is a tensor with four indices indicating the occupied orbitals $a,b$ and the virtual ones $r,s$.
$U_{int}$ is designed to prepare the ERIs from the ground state $|0\rangle$,
\begin{equation}
    U_{int}|0\rangle=\frac{\gamma_{abrs}}{\sum_{abrs}|\gamma_{abrs}|^2}|abrs\rangle
    \label{eq_u_int_ideal}
\end{equation}
where we denote the ERI in MOs as 
\begin{equation}
    \gamma_{abrs}=
    \langle ab|rs\rangle
\end{equation}
$\gamma_{abrs}$ is introduced not only for simplicity, but also to eliminate ambiguity between the ERI tensor $\langle ab|rs\rangle$ and the quantum state $|abrs\rangle$ or $|rs\rangle$.

\begin{figure}[ht]
    \centering
    \includegraphics[width=0.85\textwidth]{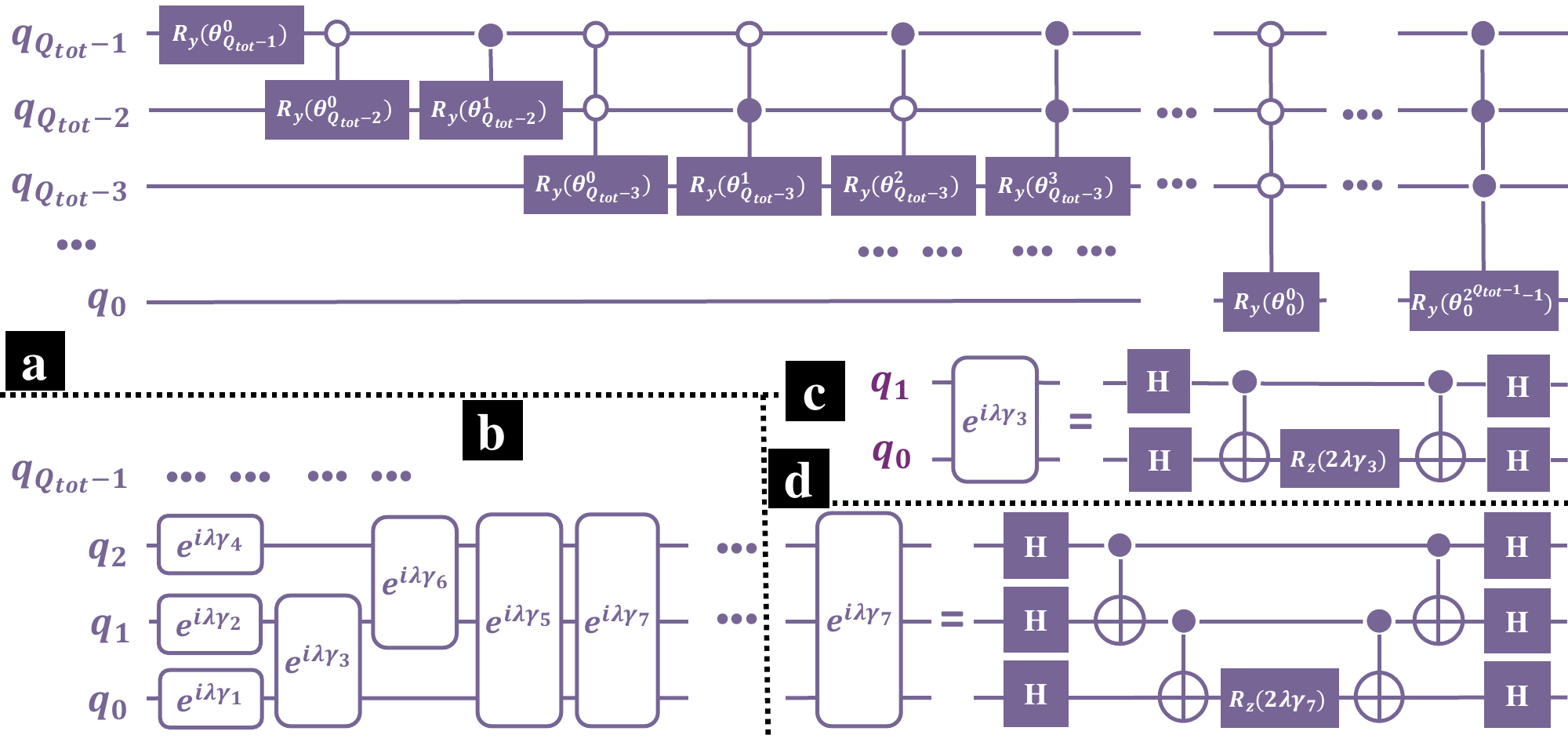}
    \caption{
    {\bf Schematic structure of the quantum circuits estimating ERI.}
    (a)The implementation of $U'_{int}$.
    There are $Q_{tot}$ qubits representing the orbitals.
    $U'_{int}$ contains a series of multi-controller $R_y$ gates with a single $R_y$ gate.
    (b)The schematic structure of $U_{int}$.
    For simplicity, only the operations applied on $q_{0,1,2}$ are depicted.
    (c)Decomposition of $e^{i\lambda\gamma_2}$, which strictly, represents $e^{i\lambda\gamma_2\sigma_x(q_0)\otimes\sigma_x(q_1)}$.
    (d)Decomposition of $e^{i\lambda\gamma_7}$, where the three qubits are $q_{0,1,2}$, the strict description is $e^{i\lambda\gamma_7\sigma_x(q_0)\otimes\sigma_x(q_1)\otimes\sigma_x(q_2)}$.
    }
    \label{figs_eri1}
\end{figure}

Theoretically, for arbitrary non-trivial ERIs there always exist $U_{int}$ that strictly guarantees Eq.(\ref{eq_u_int_ideal}).
Here we propose a general design $U'_{int}$ strictly satisfying Eq.(\ref{eq_u_int_ideal}), where we use notation $U'_{int}$ to avoid ambiguity with the improved implementation $U_{int}$, which will be discussed later in this subsection.
In Fig.(\ref{figs_eri1}a) the schematic structure of $U'_{int}$ is depicted.
Mathematically, we have
\begin{equation}
    U'_{int}=
    \left\{
    \prod_{j=1}^{Q_{tot}-1}{
    \left[
    \sum_{x=0}^{2^j-1}
    |x\rangle\langle x|\otimes
    R_y\left(
    \theta^x_{Q_{tot}-1-j}
    \right)
    %+
    %\left(
    %I^{\otimes j}
    %-%\sum_{x=0}^{2^j-1}
    %|x\rangle\langle x|
    %\right)
    %\otimes
    %I
    \right]
    }
    \right\}
    \otimes
    \left[
    R_y\left(
    \theta^0_{Q_{tot}-1}
    \right)
    \otimes I^{\otimes (Q_{tot}-1)}
    \right]
    \label{eq_uint_p}
\end{equation}
Differ from Eq.(\ref{eq_ue_improve}), the order of factors in Eq.(\ref{eq_uint_p}) is crucial.
Here each factor is set on the left side of the former ones, ensuring that the last factor ($j=Q_{tot}-1$) in the product represents the last operator (from left) in Fig.(\ref{figs_eri1}a). 
There are $Q_{tot}$ qubits representing the orbitals.
$U'_{int}$ contains a series of multi-controller $R_y$ gates with a single $R_y$ gate.

At the beginning, a simple $R_y$ gate is applied on the qubit representing the highest digit $q_{Q_{tot}-1}$.
This simple $R_y$ gate is designed to prepare the first digit of the quantum state
Recalling that $Q_{tot}=\lceil \log_2N_{tot}\rceil$, we have
\begin{equation}
    R_y(\theta^0_{Q_{tot}-1})
    =
    \sqrt{\frac{\sum_{k=0}^{2^{Q_{tot}-1}-1}|\gamma_k|^2}{\sum_{k=0}^{2^{Q_{tot}}-1}|\gamma_k|^2}}
    |0\rangle
    +\sqrt{\frac{\sum_{k'=2^{Q_{tot}-1}}^{2^{Q_{tot}}-1}|\gamma_{k'}|^2}
    {\sum_{k=0}^{2^{Q_{tot}}-1}|\gamma_k|^2}}
    |1\rangle
\end{equation}
where $k$ and $k'$ are integers representing the orbital indices $abrs$.
If we write down $k$ and $k'$ in the binary forms, then in the first sum, we have $k=(0\cdots)_{bin}$, whereas in the second sum, we have $k'=(1\cdots)_{bin}$.
In other words, we have
\begin{equation}
    \theta^0_{Q_{tot}-1}
    =
    2\arcsin\left(
    \sqrt{\frac{\sum_{k'=2^{Q_{tot}-1}}^{2^{Q_{tot}}-1}|\gamma_{k'}|^2}
    {\sum_{k=0}^{2^{Q_{tot}}-1}|\gamma_k|^2}}
    \right)
\end{equation}
Such a design guarantees that the probability to find $q_{Q_{tot}-1}$ at $|1\rangle$ or $|0\rangle$ is same to the sum of the corresponding $|\gamma|^2$ values.

The succeeding operations correspond to the first factor in the product, with $j=1$.
In Eq.(\ref{eq_uint_p}), $|x\rangle\langle x|\otimes R_y$ represents a $C^jR_y$ gate, where $|x\rangle$ is the quantum state of the $j$ control qubits, $q_{Q_{tot}-1},\cdots, q_{Q_{tot}-j}$.
As $j=1$, the summation over $x$ contains two terms, $x=0,1$.
The two $CR_y$ gates are designed to prepare quantum states corresponding to the first two digits, ensuring that
\begin{equation}
    R_y(\theta^0_{Q_{tot}-2})|0\rangle
    =
    \sqrt{\frac{\sum_{k=0}^{2^{Q_{tot}-2}-1}|\gamma_k|^2}
    {\sum_{k=0}^{2^{Q_{tot}-1}-1}|\gamma_k|^2}}
    |0\rangle
    +\sqrt{\frac{\sum_{k'=2^{Q_{tot}-2}}^{2^{Q_{tot}-1}-1}|\gamma_{k'}|^2}
    {\sum_{k=0}^{2^{Q_{tot}-1}-1}|\gamma_{k'}|^2}}
    |1\rangle
\end{equation}
and
\begin{equation}
    R_y(\theta^1_{Q_{tot}-2})|0\rangle
    =
    \sqrt{\frac{\sum_{k=2^{Q_{tot}-1}}^{2^{Q_{tot}-1}+2^{Q_{tot}-2}-1}|\gamma_k|^2}
    {\sum_{k=2^{Q_{tot}-1}}^{2^{Q_{tot}}-1}|\gamma_{k}|^2}}
    |0\rangle
    +\sqrt{\frac{\sum_{k'=2^{Q_{tot}-1}+2^{Q_{tot}-2}}^{2^{Q_{tot}}-1}|\gamma_{k'}|^2}
    {\sum_{k=2^{Q_{tot}-1}}^{2^{Q_{tot}}-1}|\gamma_{k'}|^2}}
    |1\rangle
\end{equation}
Similarly, the other multi-controller gates can be designed with
\begin{equation}
    R_y(\theta^x_{Q_{tot}-1-j})|0\rangle
    =
    \sqrt{
    \frac{\sum_{k=2^{Q_{tot}-j}x}^{2^{Q_{tot}-j-1}(2x+1)-1}|\gamma_x|^2}
    {\sum_{k=2^{Q_{tot}-j}x}^{2^{Q_{tot}-j}(x+1)-1}|\gamma_x|^2}}
    |0\rangle
    +
    \sqrt{
    \frac{\sum_{k=2^{Q_{tot}-j-1}(2x+1)}^{2^{Q_{tot}-j}(x+1)-1}|\gamma_x|^2}
    {\sum_{k=2^{Q_{tot}-j}x}^{2^{Q_{tot}-j}(x+1)-1}|\gamma_x|^2}
    }
    |1\rangle
    \label{eq_u_int_theta}
\end{equation}
Notice that Eq.(\ref{eq_u_int_theta}) is valid only if $\sum_{k=2^{Q_{tot}-j}x}^{2^{Q_{tot}-j}(x+1)-1}|\gamma_x|^2>0$, otherwise $R_y(\theta^x_{Q_{tot}-1-j})$ changes into an identical gate.

Though $U_{int}'$ strictly satisfy Eq.(\ref{eq_u_int_ideal}), it is not efficient to estimate the ERI with $U'_{int}$ on NISQ devices, especially when $N_{tot}$ is a large number.
As indicated in Eq.(\ref{eq_uint_p}), there are $2^j$ $C^jR_y$ gates in the $j-th$ factor.
Recalling that it takes $\mathcal{O}(k^2)$ CNOT gates and single qubit gates to prepare a $C^kR_y$ gate.
The time complexity of $U'_{int}$ is
\begin{equation}
    \mathcal{O}\left(
    \sum_{j=1}^{Q_{tot}-1}j^22^j
    \right)
\end{equation}
and
\begin{equation}
    \mathcal{O}\left({Q^2_{tot}2^{Q_{tot}}}\right)
    \leq
    \mathcal{O}\left(
    \sum_{j=1}^{Q_{tot}-1}j^22^j
    \right)
    \leq
    \mathcal{O}\left(Q^3_{tot}2^{Q_{tot}}\right)
\end{equation}
Briefly, time complexity of $U'_{int}$ implementation is exponential to $Q_{tot}$, and polynomial to $N_{tot}$.
Notice that the time spent to figure out $\theta$ values are not taken into account, the overall time consuming to $U_{int}'$ can be more `expensive' in experiments.

Therefore, we have to consider a `trade-off' between the efficiency and accuracy.
Comparing to $U'_{int}$, the new implementation should be less time-consuming.
In other words, we need an implementation with a shallower circuit, with parameters that can be obtained more easily. 
Here, we propose such an implementation denoted as $U_{int}$.
Intuitively, a sequence of Pauli-x gates can directly convert the ground state into the quantum states representing certain orbitals $|x\rangle$, as
\begin{equation}
    \bigotimes_{j=0}^{Q_{tot}}
    \sigma_{x_j}|0\rangle
    =|x\rangle
    \label{eq_sigmax_sq}
\end{equation}
where $x$ is a non negative integer representing the orbital $a,b,r,s$, whereas $x_j$ is the $j-th$ digit in the binary form of $x$.
$\sigma$ indicate Pauli matrices, $\sigma_0$ represents the identity gate and $\sigma_1$ represents the Pauli-X gate.
A single sequence in Eq.(\ref{eq_sigmax_sq}) is unitary, yet the combination of many is not.
Therefore, we instead consider the exponential form
\begin{equation}
    U^{ideal}_{int}(\lambda)=
    \exp{\left\{
    i\lambda
    \sum_{x=0}^{N_{tot}-1}
    \left(
    \gamma_{x}
    \bigotimes_{j=0}^{Q_{tot}}
    \sigma_{x_j}
    \right)
    \right\}}
    \label{eq_u_int_exp1}
\end{equation}
where $0<\lambda|\gamma_x|\ll 1$.
The superscript $ideal$ indicates that Eq.(\ref{eq_u_int_exp1}) is an ideal description.
Though the exponential form itself is an approximation to the `ideal' operation as expected, Eq.(\ref{eq_u_int_exp1}) itself can hardly be implemented on a real machine perfectly.
Applying the first-order Trotter decomposition\cite{lloyd1996universal}, we intend to approximate Eq.(\ref{eq_u_int_exp1}) with
\begin{equation}
    U_{int}(\lambda)=
    \prod_{x=0}^{N_{tot}-1}
    \left\{
    \exp{
    \left(
    i\lambda
    \gamma_{x}
    \bigotimes_{j=0}^{Q_{tot}}
    \sigma_{x_j}
    \right)
    }
    \right\}
    \label{eq_u_int}
\end{equation}
In Fig.(\ref{figs_eri1}b) the schematic implementation of $U_{int}$ is demonstrated, where for simplicity, only the operations acted on $q_0,q_1,q_2$ are plotted, corresponding to the first 7 $\gamma$ values.
Notation $e^{i\lambda\gamma_x}$ in Fig.(\ref{figs_eri1}b) indicates $\exp
    {\left(
    i\lambda
    \gamma_{x}
    \bigotimes_{j=0}^{Q_{tot}}
    \sigma_{x_j}
    \right)}$.
As a simple instance, binary form of $x=3$ is $0\cdots 011$.
The implementation of interaction $\exp(i\lambda\gamma_3\sigma_x(q_1)\otimes\sigma_x(q_0))$ is demonstrated as $Fig.(\ref{figs_eri1})c$, where $\sigma_x(q_1)$ represent a Pauli-X gate acting on $q_1$.
Similarly, in Fig.(\ref{figs_eri1}d) we demonstrate the decomposition of the $x=7$ case, which refers to $\exp(i\lambda\gamma_3\sigma_x(q_2)\otimes\sigma_x(q_1)\otimes\sigma_x(q_0))$.
More details about the decomposition of can be found in\cite{ortiz2001quantum, whitfield2011simulation}.

In fact, $U_{int}$ does not satisfy our original goal Eq.(\ref{eq_u_int_ideal}).
$U_{int}(\lambda)$ instead prepares an approximation of the ERI, for $\lambda|\gamma_x|\ll 1$,
\begin{equation}
    U_{int}(\lambda)|0\rangle
    =
    %\left(
    %1-i\lambda\sum_{x}\gamma_x
    %\right)
    |0\rangle
    +i\lambda\sum_{x=1}^{N_{tot}}\gamma_x|x\rangle
    +\mathcal{O}(\lambda^2)
    \label{eq_u_int_approx}
\end{equation}
Though $U_{int}$ prepares an approximation, instead of an accurate estimation of the ERI, it leads to considerable advantages against $U_{int}'$, particularly in feasibility and efficiency.
In $U_{int}'$ multi controller $R_y$ gates are harnessed to prepare the ERI, while in $U_{int}$,  $\exp
    {\left(
    i\lambda
    \gamma_{x}
    \bigotimes_{j=0}^{Q_{tot}}
    \sigma_{x_j}
    \right)}$
is employed.
A $C^kR_y$ gate can be decomposed into $\mathcal{O}(k^2)$ CNOT gates or single qubit gates, whereas $\exp
    {\left(
    i\lambda
    \gamma_{x}
    \bigotimes_{j=0}^{Q_{tot}}
    \sigma_{x_j}
    \right)}$
with $k$ digit as $1$ contains only $\mathcal{O}(k)$ CNOT gates or single qubit gates.
Overall, the time complexity of $U_{int}$ is
\begin{equation}
    \mathcal{O}(N_{tot}Q_{tot}) = \mathcal{O}(Q_{tot}2^{Q_{tot}}) = \mathcal{O}(N_{tot}\log_2N_{tot})
\end{equation}
thereby $U_{int}$ is much shallower than $U_{int}'$.
Moreover, it is much easier to implement the $\sigma_x$ sequence on real machines, comparing to the multi-controller $R_y$ gates.
More discussion about the feasibility of $U_{int}$ are presented in the following sections.

By the end of this subsection, we would like to present a trick in the design of $U_{int}$.
The ERI tensor is often sparse, and an alternative denoted as $U^{alter}_{int}$ might be a better choice in some certain cases.
The definition of $U^{alter}_{int}$ in first order Trotter decomposition is
\begin{equation}
    U^{alter}_{int}(\lambda)=
    \prod_{x=0}^{N_{tot}-1}
    \left\{
    \exp{
    \left(
    i\lambda
    \gamma_{x}
    \bigotimes_{j=0}^{Q_{tot}}
    \sigma_{x_j\oplus y_j}
    \right)
    }
    \right\}
    \cdot
    \bigotimes_{j=0}^{Q_{tot}}
    \sigma_{y_j}
    \label{eq_uint_alter}
\end{equation}
$U_{int}^{alter}$ firstly converts the $|0\rangle$ into a certain state $|y\rangle$, then the Pauli-X sequence is applied, similarly to $U_{int}$.
Sometimes the alternative design can further reduce the requirement in connectivity and number of element gates in decomposition.
$U_{int}^{alter}$ is equivalent to $U_{INT}$ in the main article.

\subsection{Circuits transforming AO to MO}

According to Eq.(\ref{expansion_phi}), MOs can be written as a linear combination of the AOs, where the expansion coefficients $c_{ka}$ are derived from the Hartree-Fock calculations.
Though the MOs are orthogonal orbitals, the AOs are often not.
As an instance, the typical GTOs\cite{boys1950electronic} are not an orthogonal set.
Thus, it is extremely challenging to employ simple unitary operations to transform AOs to MOs directly.
However, in the MP2 calculations, we do not need to transform AOs to MOs one by one.
Instead, we would like to get the ERI in MO basis, where the ERI in AO basis is often obtained in the Hartree-Fock calculations.
In this section, we propose a circuit design denoted as $U_{trans}$, which is employed to calculate the ERI in MO with given ERI in AO.
For simplicity, here we map the AOs to the computational eigenstates, such as $|klmn\rangle$.
Then $U_{trans}$ guarantees that
\begin{equation}
    U_{trans}(|klmn\rangle\otimes|0\rangle)
    =
    |klmn\rangle\otimes
    \left(
    \sum_a\sum_b\sum_r\sum_s
    c^*_{ka}c^*_{lb}c_{mr}c_{ns}
    |abrs\rangle
    \right)
    \label{eq_u_trans_aim}
\end{equation}
where the coefficients $c$ are the same ones as in Eq.(\ref{eq_expansion_eri}).
There are two registers, one representing the AOs, denoted as $q^{AO}$, whereas the ones for Mos are denoted as $q^{MO}$.
The qubits $q^{MO}$ are initialized at ground state $|0\rangle$.
$q^{AO}$ are employed as control qubits, converting the target $q^{MO}$ to the corresponding superposition.

Before diving deep to the implementation, we would like to firstly illustrate the function of $U_{trans}$.
$U_{trans}$ is designed to prepare the ERI in MOs from a given quantum state representing ERI in AOs.
As demonstrated in Sec.(\ref{Circuits estimating ERI}), $U'_{int}$ or $U_{int}(\lambda)$ are designed to prepare the quantum state representing ERI in MO, and these operations can as well prepare the quantum state representing ERI in AO.
Consider the $U'_{int}$ preparing the ERI in AOs $\langle kl|mn\rangle$, and denote 
\begin{equation}
    \gamma_{klmn}=\frac{\langle kl|mn\rangle}
    {\sum_{klmn}|\langle kl|mn\rangle|^2}
\end{equation}
then we have
\begin{equation}
    \begin{split}
    U_{trans}
    \left[\left(U'_{int}|0\rangle_{AO}\right)\otimes|0\rangle_{MO}\right]
    &=U_{trans}
    \left[\left(\sum_{klmn}\gamma_{klmn}|klmn\rangle\right)\otimes|0\rangle\right]
    \\
    &=\sum_{klmn}\gamma_{klmn}\left[
    U_{trans}\left(|klmn\rangle\otimes|0\rangle\right)
    \right]
    \\
    &=
    \sum_{klmn}\gamma_{klmn}\left[
    |klmn\rangle\otimes
    \left(
    \sum_{abrs}c^*_{ka}c^*_{lb}c_{mr}c_{ns}
    |abrs\rangle
    \right)
    \right]
    \\
    &=
    \sum_{abrs}
    \left[
    \left(
    \sum_{klmn}
    \gamma_{klmn}c^*_{ka}c^*_{lb}c_{mr}c_{ns}|klmn\rangle
    \right)
    \otimes
    |abrs\rangle
    \right]
    \end{split}
    \label{eq_math_u_trans}
\end{equation}
Then we measure $q^{MO}$, theoretically, the probability to get result $|abrs\rangle$ is
\begin{equation}
    \left|\sum_{klmn}
    \gamma_{klmn}c^*_{ka}c^*_{lb}c_{mr}c_{ns}\right|^2
    =
    \left|
    \frac{1}{\sum_{klmn}|\langle kl|mn\rangle|^2}
    c^*_{ka}c^*_{lb}c_{mr}c_{ns}
    \langle kl|mn\rangle
    \right|^2
\end{equation}
Notice that
\begin{equation}
    \langle ab|rs\rangle=
    \sum_{klmn}c^*_{ka}c^*_{lb}c_{mr}c_{ns}\langle kl|mn\rangle
\end{equation}
Eq.(\ref{eq_math_u_trans}) provides us with an approach to estimate the ERI in MOs from given quantum state representing the ERI in AOs.
In Eq.(\ref{eq_math_u_trans}) we only consider the $U'_{int}$ for simplicity.
Similarly, $U_{int}(\lambda)$ also works, as
\begin{equation}
\begin{split}
    &U_{trans}
    \left[\left(U_{int}(\lambda)|0\rangle_{AO}\right)\otimes|0\rangle_{MO}\right]
    \\
    =&U_{trans}
    \left[\left(|0\rangle +i\lambda\sum_{klmn}\gamma_{klmn}|klmn\rangle
    +\mathcal{O}(\lambda^2)\right)
    \otimes|0\rangle\right]
    \\
    =&
    U_{trans}(|0\rangle\otimes|0\rangle)
    +i\lambda\sum_{klmn}\left(
    \gamma_{klmn}U_{trans}|klmn\rangle\otimes|0\rangle\right)
    +\mathcal{O}(\lambda^2)
    \\
    =&
    U_{trans}(|0\rangle\otimes|0\rangle)+
    i\lambda\sum_{klmn}\gamma_{klmn}\left[
    |klmn\rangle\otimes
    \left(
    \sum_{abrs}c^*_{ka}c^*_{lb}c_{mr}c_{ns}
    |abrs\rangle
    \right)
    \right]
    +\mathcal{O}(\lambda^2)
    \\
    =&
    U_{trans}(|0\rangle\otimes|0\rangle)+
    i\lambda\sum_{abrs}
    \left[
    \left(
    \sum_{klmn}
    \gamma_{klmn}c^*_{ka}c^*_{lb}c_{mr}c_{ns}|klmn\rangle
    \right)
    \otimes
    |abrs\rangle
    \right]
    +\mathcal{O}(\lambda^2)
    \end{split}
    \label{eq_math_u_trans2}
\end{equation}
Here we need to ensure that $|0\rangle$ does not correspond to any valid orbitals in both AOs and MOs, and $U_{trans}|0\rangle=|0\rangle$.
If so, $U_{trans}$ is able to prepare an approximation of the ERI in MOs, from the given quantum state representing approximation of the ERI in AOs.
According to Eq.(\ref{eq_math_u_trans2}), if we only measure $q^{MO}$ after the operations, then theoretically, the probability to find $q^{MO}$ at state $|abrs\rangle$ is 
\begin{equation}
    \left|i\lambda\sum_{klmn}
    \gamma_{klmn}c^*_{ka}c^*_{lb}c_{mr}c_{ns}
    +\mathcal{O}(\lambda^2)
    \right|^2
    =
    \lambda^2\left|
    \frac{1}{\sum_{klmn}|\langle kl|mn\rangle|^2}
    c^*_{ka}c^*_{lb}c_{mr}c_{ns}
    \langle kl|mn\rangle
    \right|^2+\mathcal{O}(\lambda^3)
    \label{eq_u_trans_output}
\end{equation}
For small $\lambda$ values, the probability in Eq.(\ref{eq_u_trans_output}) is proportional to $\lambda^2$.
Thereby, the ERI in MO can be estimated via a linear regression process with a range of $\lambda^2$.

The function of $U_{trans}$ is demonstrated as above, thereafter we will focus on the implementation.
The fundamental feature of $U_{trans}$ is as described in Eq.(\ref{eq_u_trans_aim}), where the indices $a,b,r,s$ and $k,l,m,n$ represent independent occupied or virtual orbitals in MOs and AOs.
The transformation does not mix up these orbitals.
Instead, the first index $a$ as MO only corresponds to the index $k$ in AO, and so do the others.
Thereby, it is reasonable to separate $U_{trans}$ into four parts, denoted as $U_{trans}^{occ}$ for indices $a,b,k,l$ and $U_{trans}^{vir}$ for $r,s,m,n$.
The superscripts indicate the occupied or virtual orbitals.

\begin{figure}[ht]
    \centering
    \includegraphics[width=0.85\textwidth]{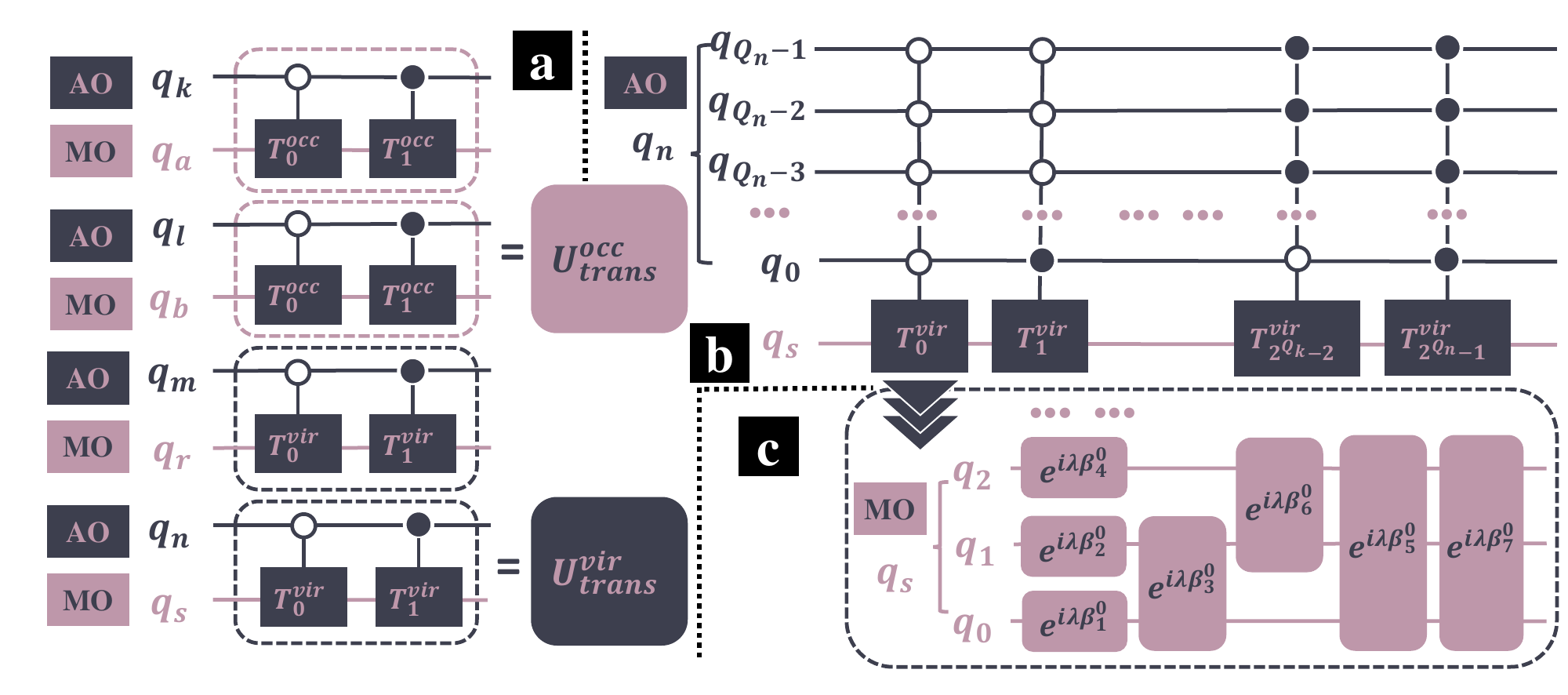}
    \caption{
    {\bf Schematic structure of the quantum circuits transforming AOs to MOs.}
    (a)(Left)The implementation of $U_{trans}$.
    Qubits representing AOs are depicted in pink, whereas qubits for MOs are in deep navy.
    $q_a$, $q_b$, $q_k$, $q_l$ are occupied orbitals, corresponding to $U_{trans}^{occ}$, while the others are virtual, corresponding to $U_{trans}^{vir}$.
    (b)(Right upper)The schematic structure of $U_{trans}^{vir}$.
    (c)(Right bottom)The schematic structure of $T_{0}^{vir}$, where only three qubits in $q_s$ are plotted for simplicity.
    The design is similar to $U_{eri}$ as depicted in Fig.(\ref{figs_eri1}).
    }
    \label{figs_utrans}
\end{figure}

The schematic structure of $U_{trans}$ is depicted in Fig.(\ref{figs_utrans}a).
Qubits representing AOs are depicted in pink, whereas qubits for MOs are in deep navy.
$q_a$, $q_b$, $q_k$, $q_l$ are occupied orbitals, corresponding to $U_{trans}^{occ}$, while the others are virtual, corresponding to $U_{trans}^{vir}$.
For simplicity, both the control and target qubits are depicted as one single qubit in Fig.(\ref{figs_utrans}a).
In the mathematical manner, $U_{trans}^{occ}$ is designed as
\begin{equation}
    U_{trans}^{occ}=
    \bigotimes_{j=0}^{2^{Q_{occ}}-1}
    \left(
    |j\rangle\langle j|\otimes
    T^{occ}_j
    \right)
\end{equation}
and 
\begin{equation}
    U_{trans}^{occ}=
    \bigotimes_{j=0}^{2^{Q_{vir}}-1}
    \left(
    |j\rangle\langle j|\otimes
    T^{vir}_j
    \right)
\end{equation}
where $Q_{occ}$ is number of qubits representing occupied orbitals.
Denote the number of occupied orbitals as $N_{occ}$, we have $Q_{occ}=\lceil\log_2N_{occ}\rceil$.
For $N_{occ}-1\leq j\leq 2^{Q_{occ}-1}$, we have $T^{occ}_j=I$ is the identical operation, and same for the virtual ones.

The schematic structure of $U_{trans}^{vir}$ is illustrated in Fig.(\ref{figs_utrans}b).
Structure of $T$ is quite similar to $U_{int}$, and we define
\begin{equation}
    T^{vir}_{a}(\lambda)
    =\prod_{x=0}^{N_{vir-1}}
    \left\{
    \exp\left(
    i\lambda\beta^a_x\bigotimes_{b=0}^{Q_{vir}-1}\sigma_{x_b}
    \right)
    \right\}
\end{equation}
where the coefficients $\beta_x^a=c_{x,a+N_{occ}}$, as the first $N_{occ}$ columns of $c$ represents the occupied orbitals.
As for $T^{occ}$, we can also define
\begin{equation}
    T^{occ}_{a}(\lambda)
    =\prod_{x=0}^{N_{vir-1}}
    \left\{
    \exp\left(
    i\lambda c_{xa}\bigotimes_{b=0}^{Q_{vir}-1}\sigma_{x_b}
    \right)
    \right\}
\end{equation}
    
Generally, there are more virtual orbitals than the occupied ones.
Thereby $T^{vir}$ is often deeper than $T^{occ}$.
In some special cases with very few occupied orbitals, it might be no more necessary to keep the approximation with Trotter expansions.
Instead, the simple $C^nR_y$ gates could be a better choice (like the design of $U'_{int}$, Sec.(\ref{Circuits estimating ERI})).
In the main article, we focus on the MP2 calculations of the simple Helium atom.
There are only 2 electrons, thereby only the $1s$ orbital is occupied at ground state.
Thereby, in our study of Helium, $T^{occ}$ is replaced with a simple $R_y$ gate.

\subsection{Circuits estimating the difference}

In this subsection, we will focus on the circuit that estimate the difference, which can be harnessed to estimate the difference of ERI, see $\langle ab|rs\rangle-\langle ab|sr\rangle$.
Moreover, it can also be employed to improve the accuracy of the exponential terms in Trotter expansion.

\begin{figure}[ht]
    \centering
    \includegraphics[width=0.85\textwidth]{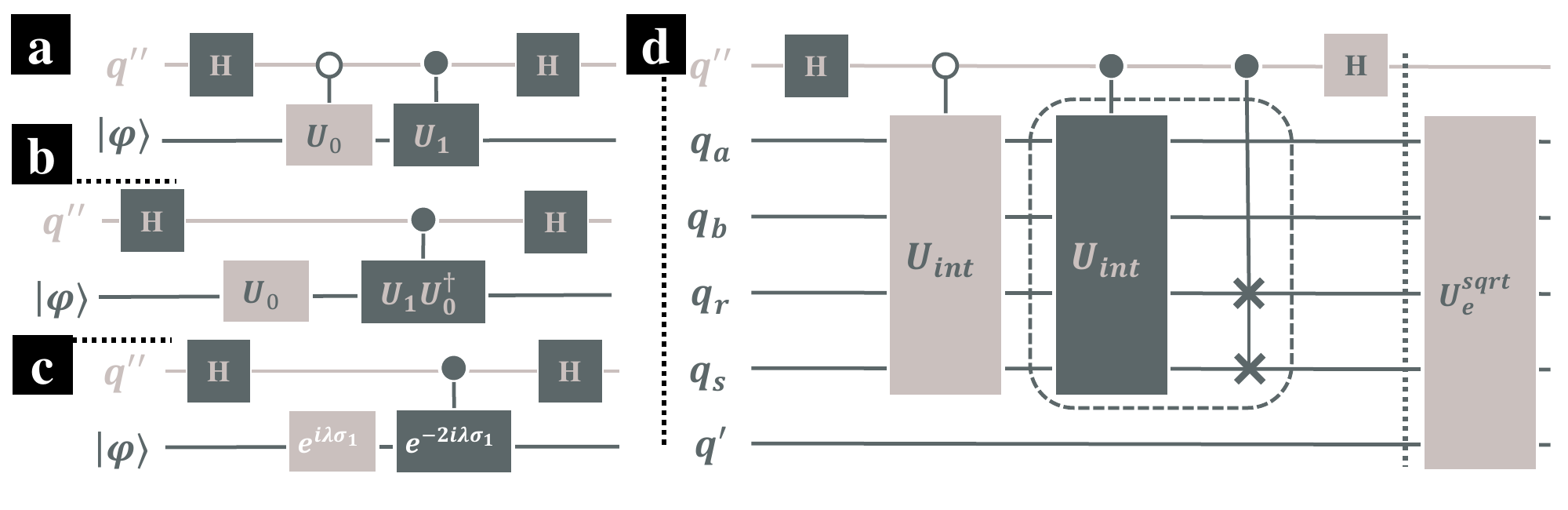}
    \caption{
    {\bf Schematic structure of the quantum circuits estimating the difference.}
    (a)Depiction of a simple circuit estimating the difference.
    $q''$ is an ancilla qubit initialized at state $|0\rangle$.
    (b)An equivalent structure of (a), where there are less $CU$ gates required.
    (c)Application on the first order Trotter decomposition.
    (d)Application on ERI estimation.
    Operations in the dashed box correspond to $U_1$ in (a), whereas $U_{int}$ on the left (colored in light gray) corresponds to $U_0$ in (a).
    All qubits are initialized at state $|0\rangle$.
    }
    \label{figs_diff}
\end{figure}

Consider the simple circuit as depicted in Fig.(\ref{figs_diff}a), the ancilla qubit is initialized as $|0\rangle$.
The output is 
\begin{equation}
    \begin{split}
        (H\otimes I)\frac{1}{\sqrt{2}}
    \left(|0\rangle\otimes U_0|\varphi\rangle
    +|1\rangle\otimes U_1|\varphi\rangle\right)
    &=
    \frac{1}{2}\left\{
    (|0\rangle+|1\rangle)\otimes U_0|\varphi\rangle
    +(|0\rangle-|1\rangle)\otimes U_1|\varphi\rangle
    \right\}
    \\
    &=
    \frac{1}{2}
    \left\{
    |0\rangle\otimes(U_0|\varphi\rangle+U_1|\varphi\rangle)
    +|1\rangle\otimes(U_0|\varphi\rangle-U_1|\varphi\rangle)
    \right\}
    \end{split}
\end{equation}
Then we measure all of the qubits, theoretically, the probability to find the ancilla qubit at $|0\rangle$ along with the others at state $|n\rangle$ is
\begin{equation}
    \frac{1}{4}\left|
    \langle n|U_0-U_1|\varphi\rangle
    \right|^2
    \label{eq_u_diff_basic}
\end{equation}
from which we are able to estimate $\langle n|U_0-U_1|\varphi\rangle$.

In Fig.(\ref{figs_diff}b) an equivalent circuit is depicted, where there contains only one $CU$ operation.
(The equivalence is due to that $U_1U_0^{\dagger} U_0=U_1$.)
The design in Fig.(\ref{figs_diff}b) is often beneficial when dealing with the exponential terms.
Consider the case where $U_0=\exp(i\lambda V)$, and $U_1=\exp(-i\lambda V)$, the probability in Eq.(\ref{eq_u_diff_basic}) is now
\begin{equation}
    \begin{split}
        \frac{1}{4}\left|
    \langle n|\exp(i\lambda V)-\exp(-i\lambda V)|\varphi\rangle
    \right|^2
    &=
    \frac{1}{4}
    \left|
    \langle n|
    i\lambda V-\frac{1}{2}\lambda^2 V^2
    -\left(-i\lambda V-\frac{1}{2}\lambda^2 V^2\right)
    +\mathcal{O}(\lambda^3)
    |\varphi\rangle
    \right|^2
    \\
    &=
    \frac{1}{2}
    \left|
    \langle n|2i\lambda V+\mathcal{O}(\lambda^3)|\varphi\rangle
    \right|^2
    \end{split}
\end{equation}
Thereby the $\lambda^2V^2$ terms are eliminated.
In Fig.(\ref{figs_diff}c) we depict an example where $V=\sigma_1$ is the Pauli-X gate.
In our recent work the same circuits are employed for more accurate estimation of perturbation\cite{li2023toward}.

Furthermore, such circuit estimating difference is fundamental in the MP2 calculations, and is applied to estimate the antisymmerized two electron integrals $\langle ab||rs\rangle$ as defined in Eq.(\ref{eq_anti}).
As discussed in Sec.(\ref{Circuits estimating ERI}), $U_{int}$ is designed to prepare the ERI $\langle ab|rs\rangle$ (or the approximation) in a quantum state.
Schematic structure of circuit estimating $\langle ab||rs\rangle$ is depicted in Fig.(\ref{figs_diff}d). 
All qubits are initialized at $|0\rangle$ at the beginning.
$U_{int}$ on left (colored in light grey) is harnessed as $U_0$, whereas the operations in the dashed box are employed as $U_1$, generating the $\langle ab|sr\rangle$ components.
The connected cross symbols represent the SWAP gate between $q_r$ and $q_s$, and here there is a control-SWAP gate.
The SWAP gate $SWAP_{r,s}$ makes,
\begin{equation}
    SWAP_{r,s}|abrs\rangle=|absr\rangle
\end{equation}
Recalling Eq.(\ref{eq_u_int_ideal}), we have
\begin{equation}
    SWAP_{r,s}U_{int}|0\rangle=\frac{\gamma_{abrs}}{\sum_{abrs}|\gamma_{abrs}|^2}|absr\rangle
\end{equation}
At the stage noted by the dashed line before $U_e$ in Fig.(\ref{figs_diff}d), the quantum state is ($q'$ is still $|0\rangle$,and not included in the formulas below)
\begin{equation}
    \begin{split}
        &\frac{1}{2}\left\{
    |0\rangle\otimes(U_0|0\rangle+U_1|0\rangle)
    +|1\rangle\otimes(U_0|0\rangle-U_1|0\rangle)
    \right\}
    \\
    =&
    \frac{1}{2\sum_{abrs}|\gamma_{abrs}|^2}
    \left\{
    |0\rangle\otimes
    \left(
    \sum_{abrs}\gamma_{abrs}|abrs\rangle+\sum_{abrs}\gamma_{abrs}|absr\rangle
    \right)
    +|1\rangle\otimes
    \left(
    \sum_{abrs}\gamma_{abrs}|abrs\rangle-\sum_{abrs}\gamma_{abrs}|absr\rangle
    \right)
    \right\}
    \\
    =&
    \frac{1}{2\sum_{abrs}|\gamma_{abrs}|^2}\sum_{abrs}
    \left\{
    |0\rangle\otimes(\gamma_{abrs}+\gamma_{absr})|abrs\rangle
    +
    |1\rangle\otimes(\gamma_{abrs}-\gamma_{absr})|abrs\rangle
    \right\}
    \end{split}
\end{equation}
Thereafter, if we measure $q_{a}, q_b, q_r,q_s$ and $q''$ at this stage, the theoretical probability to find $q''$ at $|1\rangle$, along with $q_{a}, q_b, q_r,q_s$ at state $|abrs\rangle$ is
\begin{equation}
    \frac{1}{4\left(\sum_{abrs}|\gamma_{abrs}|^2\right)^2}
    \left|\gamma_{abrs}-\gamma_{absr}\right|^2
\end{equation}
from which $\langle ab||rs\rangle$ can be estimated.
Yet our aim is not to estimate a single $\langle ab||rs\rangle$.
Instead, we are pursuing to the MP2 correlation energy as shown in Eq.(\ref{eq_mp2}).
Thus, we do not measure the qubits, but apply a succeeding operation $U_e^{sqrt}$.
Substitute Eq.(\ref{eq_u_e_sqrt}), the final quantum state is
\begin{equation}
    \begin{split}
        &U_e^{sqrt}\frac{1}{2\sum_{abrs}|\gamma_{abrs}|^2}\sum_{abrs}
    \left\{
    |0\rangle\otimes(\gamma_{abrs}+\gamma_{absr})|abrs\rangle
    +
    |1\rangle\otimes(\gamma_{abrs}-\gamma_{absr})|abrs\rangle
    \right\}
    \\
    =&
    \sum_{abrs}
    \left\{
    \frac{\gamma_{abrs}+\gamma_{absr}}{2\sum_{a'b'r's'}|\gamma_{a'b'r's'}|^2}
    |0\rangle\otimes|abrs\rangle\otimes
    \left[
    \left(1-\left|\frac{C_e^{sqrt}}{\epsilon_a+\epsilon_b-\epsilon_r-\epsilon_s}\right|\right)^{\frac{1}{2}}|0\rangle
    +
    \left(\frac{C_e^{sqrt}}{\epsilon_a+\epsilon_b-\epsilon_r-\epsilon_s}\right)^{\frac{1}{2}}|1\rangle
    \right]
    \right\}
    \\
    &+
    \sum_{abrs}
    \left\{
    \frac{\gamma_{abrs}-\gamma_{absr}}{2\sum_{a'b'r's'}|\gamma_{a'b'r's'}|^2}
    |1\rangle\otimes|abrs\rangle\otimes
    \left[
    \left(1-\left|\frac{C_e^{sqrt}}{\epsilon_a+\epsilon_b-\epsilon_r-\epsilon_s}\right|\right)^{\frac{1}{2}}|0\rangle
    +
    \left(\frac{C_e^{sqrt}}{\epsilon_a+\epsilon_b-\epsilon_r-\epsilon_s}\right)^{\frac{1}{2}}|1\rangle
    \right]
    \right\}
    \end{split}
\end{equation}
At the end, we need to measure $q''$ and $q'$.
Theoretically, the probability to find both of the qubits at state $|1\rangle$ is
\begin{equation}
    \left|\frac{C_e^{sqrt}}{2\sum_{abrs}|\gamma_{abrs}|^2}\right|^2\cdot
    \sum_{abrs}\frac{\left|\gamma_{abrs}-\gamma_{absr}\right|^2}{|\epsilon_a+\epsilon_b-\epsilon_r-\epsilon_s|}
    \label{eq_u_diff_final}
\end{equation}
Recalling the definition of $\gamma_{abrs}=\langle ab|rs\rangle$, the MP2 correlation energy finally appears in Eq.(\ref{eq_u_diff_final}). 

As a brief conclusion, in this section we thoroughly demonstrated the fundamental circuits for MP2 calculations on quantum devices.
In the succeeding sections, we will concentrate on the implementations on real machines.
Limitations of NISQ devices raises extra challenges, whereas the features of certain systems, on the contrast, considerably simplify the circuits.

\section{Implementation on Real Machines}
\label{Implementation on Real Machines}

\subsection{Implementation of $C^nR_y$ gates}
Multi qubit operations are inevitable in the implementation of $U_{e}$, $U_{int}$ and $U_{trans}$.
In $U_e$ the $C^nR_y$ gates are included, whereas in $U_{int}$, $U_{trans}$ the CNOT gate sequences are include to implement the exponential terms.
Generally, the multi controller gates are often much more demanding, which is our focus in this subsection.

\begin{figure}[ht]
    \centering
    \includegraphics[width=0.85\textwidth]{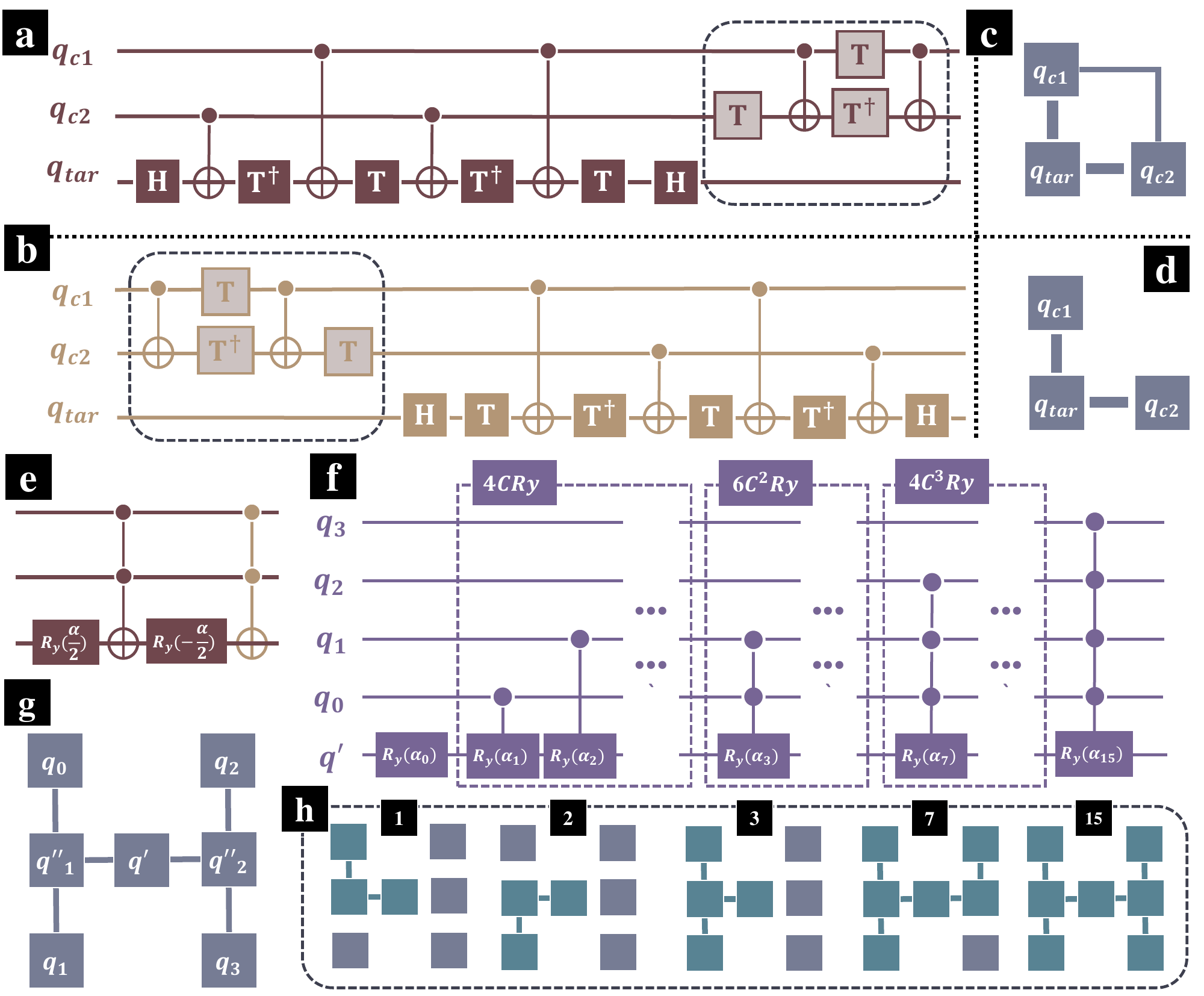}
    \caption{
    {\bf Depiction of the multi controller gates.}
    (a)Decomposition of Toffoli gate. $q_{c1}$ and $q_{c2}$ are two control qubits, whereas $q_{tar}$ is the target.
    Operations in the dashed box raises requirement for the connectivity between the two control qubits.
    (b)Decomposition of the inverse of CNOT gate, which is still a CNOT gate.
    (c)Required connectivity of CNOT gate.
    (d)Required connectivity of $C^2R_y$ gate, which is decomposed as shown in (e).
    (f)$U_e$ with four control qubits, which is same to Fig.(\ref{figs_ue1}b).
    (g)The required connectivity of $U_e$ in (f), $q''$ are ancilla qubits.
    (h)Qubits colored in azure are involved in the multi controller gates, whereas the ones colored in gray are not.
    Qubits are assigned in the `H' shape as shown in (g).
    Numbers above the qubits correspond to the operations in (f).
    As an example, `1' corresponds to $R_y(\alpha_1)$ in (f), which is a $CR_y$ gate, and $q_0$, $q''_1$, $q'$ are involved in the implementation.
    }
    \label{figs_ccnot}
\end{figure}

At the very beginning, we would like to start from Toffoli gate (also CCNOT gate, or $C^2NOT$ gate), which is one of the fundamental elements of multi-controller gates.
Decomposition of a simple Toffoli gate is depicted in Fig.(\ref{figs_ccnot}a).
$q_{c1}$ and $q_{c2}$ are two control qubits, whereas $q_{tar}$ is the target.
In addition to the CNOT gates between $q_{c}$ and $q_{tar}$, the operations in the dashed box requires connectivity between the two control qubits.
Thereby to implement such a simple Toffoli gate, there raises requirement for connectivity among the three qubits, as depicted in Fig.(\ref{figs_ccnot}c).
Yet not all devices satisfy the demandingness.
Experimentally, the SWAP gates are often employed to fulfill the implementation of Toffoli gates.
Even though, the existence of SWAP gate pairs can raise extra noises and errors.

Promisingly, the $C^nR_y$ gates in $U_e$ can further be decomposed into a $C^nNOT$ pair along with two single qubit $R_y$ gates.
In Fig.(\ref{figs_ccnot}e) the decomposition of $C^2R_y$ gate is presented.
Recalling that the Toffoli gate is an involutory gate, as its inverse is still itself.
Thereafter the Toffoli gate can also be decomposed as depicted in Fig.(\ref{figs_ccnot}b).
Substitute the decomposition in Fig.(\ref{figs_ccnot}a,b) into Fig.(\ref{figs_ccnot}e)((a) for the left one, and (b) for the right one), then operations in the dashed box all cancel out.
Thus, in the implementation of $C^2R_y$ gate, there is no two qubit gates between the control qubits, the corresponding connectivity is depicted in Fig.(\ref{figs_ccnot}d).
In our recent work\cite{li2023toward}, we applied the similar approach to build $C^2R_y$ gates.

Furthermore, we can implement $C^3R_y$ and $C^4R_y$ gates with the Toffoli gates.
Consider $U_e$ depicted in Fig.(\ref{figs_ccnot}f) (same to Fig.(\ref{figs_ue1}b)), there are branches of multi controller gates with no more than 4 control qubits.
The required connectivity is depicted in Fig.(\ref{figs_ccnot}g).
The 7 qubits are assigned as an `H' shape, where the qubits are connected if there are two qubit gates applied between them, and $q''$ are ancilla qubits.
In Fig.(\ref{figs_ccnot}h) we demonstrate the involved qubits in the multi controller gates.
Qubits colored in azure are involved, whereas the ones colored in gray are not.
Numbers above the qubits correspond to the operations in Fig.(\ref{figs_ccnot}f), whereas the squares correspond to the qubits in Fig.(\ref{figs_ccnot}g).
For example, `1' corresponds to $R_y(\alpha_1)$, which is a $CR_y$ gate in Fig.(\ref{figs_ccnot}f) with $q_0$ as the control qubit and $q'$ as target.
Notice that in in Fig.(\ref{figs_ccnot}g) $q_0$ and $q'$ are not neighbors, the ancilla qubit $q''_1$ is thus included.
Firstly apply $CNOT$ gate between $q_0$, $q''_1$.
Then implement the $CR_y$ gate on $q''_1$ and $q'$.
Thereafter another $CNOT$ gate between $q_0$, $q''_1$ is required to reset the ancilla qubit to state $|0\rangle$.
Thereby $q_0$, $q''_1$, $q'$ are involved in the implementation.

\begin{figure}[ht]
    \centering
    \includegraphics[width=0.7\textwidth]{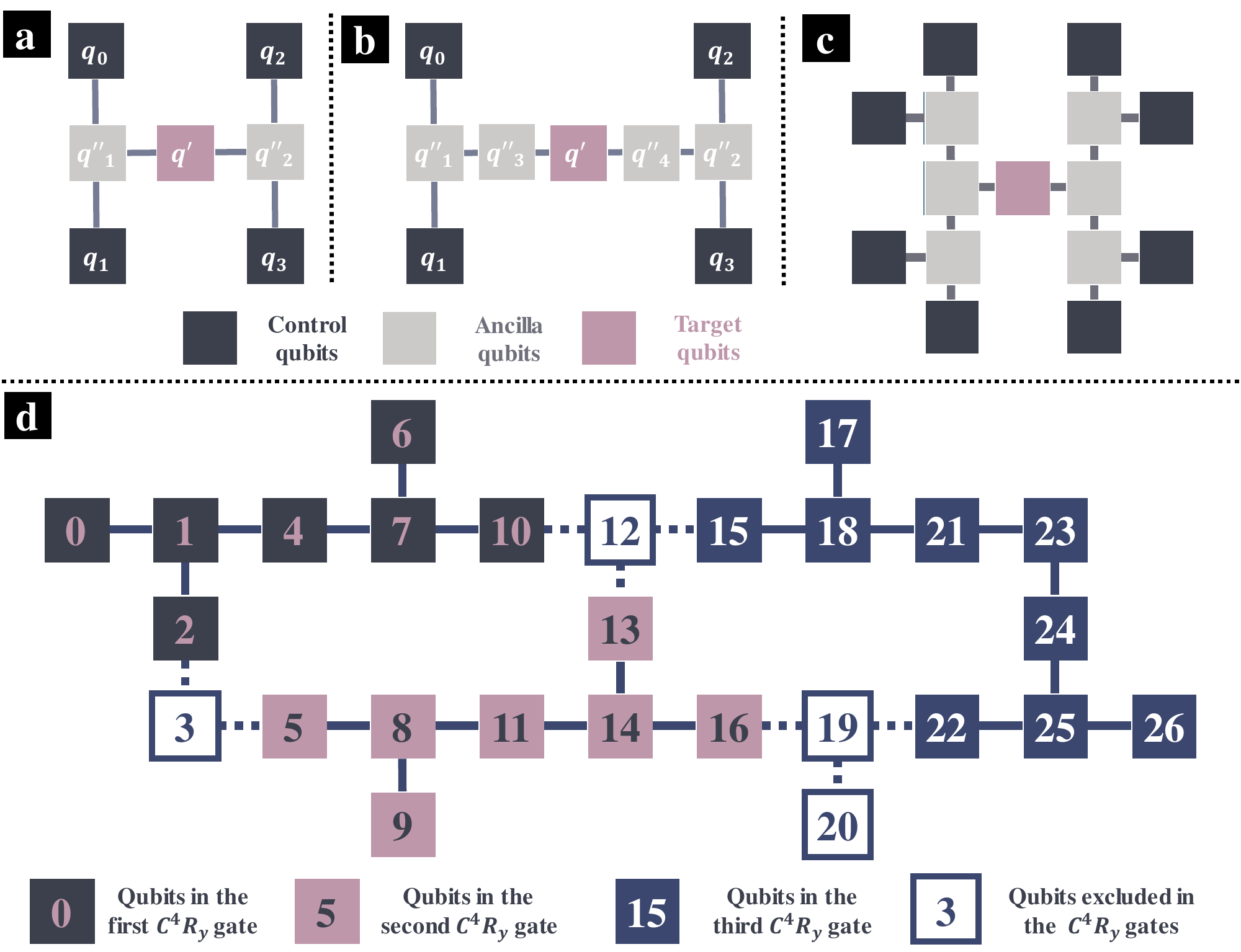}
    \caption{
    {\bf Connectivity in implementation of the $C^nR_y$ gates.}
    (a) The standard connectivity of $C^3R_y$ or $C^4R_y$ gates.
    There are CNOT gates applied on the connected qubits.
    (b) An alternative implementation, where two additional ancilla qubits $q''_{3}$, $q''_4$ are included, as $q'$ is not derictly connected to $q''_{1}$, $q''_2$.
    (c) Connectivity in the implementation fo $C^8R_y$ gates.
    In (a,b,c), control qubits are colored in deep navy, ancilla qubits are colored in light grey, and target qubit is colored in pink.
    (d) Connectivity of the IBM 27-qubit quantum chips.
    The numbers indicate the corresponding qubit on IBM quantum computers.
    The connected squares infer connected qubits on real machines.
    Three $C^4R_y$ gates can run simultaneously on the 27-qubit quantum chip.
    }
    \label{figs_para}
\end{figure}

Sometimes, the ancilla qubits are not directly connected to the target qubits on real machines.
Then we need to slightly change the implementation of $C^3R_y$ and $C^4R_y$ gates.
In Fig.(\ref{figs_para}a) the standard connectivity of $C^3R_y$ or $C^4R_y$ gates is depicted.
If the ancilla qubits are not directly connected to the target qubits, then we can apply the alternative implementation as shown in Fig.(\ref{figs_para}b), where two additional ancilla qubits $q''_{3}$, $q''_4$ are included.
The design can be extended to $C^nR_y$ gates with more control qubits.
For instance, if we regard the four control qubits in the standard $C^4R_y$ gates as ancilla qubits, and connect each of them to two control qubits, we can implement the $C^8R_y$ gate, as depicted in Fig.(\ref{figs_para}c).
In Fig.(\ref{figs_para}a,b,c), control qubits are colored in deep navy, ancilla qubits are colored in light grey, and target qubit is colored in pink.

In the main article, most circuits are implemented on IBM 27-qubit machines.
Connectivity of the IBM 27-qubit quantum chips is depicted in Fig.(\ref{figs_para}d), where the numbers indicate the corresponding qubit on IBM quantum computers.
The connected squares infer connected qubits on real machines.
Three $C^4R_y$ gates can run simultaneously on the 27-qubit quantum chip, as shown in the qubits depicted in three different colors.
The first two $C^4R_y$ gates (colored in black and pink) correspond to the standard connectivity as shown in Fig.(\ref{figs_para}a), whereas the third corresponds to the alternative one as shown in Fig.(\ref{figs_para}b).
Besides, there are 4 qubits excluded to the three $C^4R_y$ gates, as the ones depicted in white background in Fig.(\ref{figs_para}d).

\subsection{Implementation of the exponential products}
Generally, the CNOT gate sequences only requires connectivity between the neighbor qubits.
Thereby, it is often less challenging to implement the exponential map of the product of Pauli spin matrices, especially comparing with the $C^nR_y$ gates.
On the contrary, the exponential map have been well-developed and widely employed in Trotterized quantum circuits\cite{pastori2022characterization, kivlichan2020improved, han2021experimental}.
In this subsection, we will present some tricky designs in the implementation of the exponential products.

\begin{figure}[ht]
    \centering
    \includegraphics[width=0.95\textwidth]{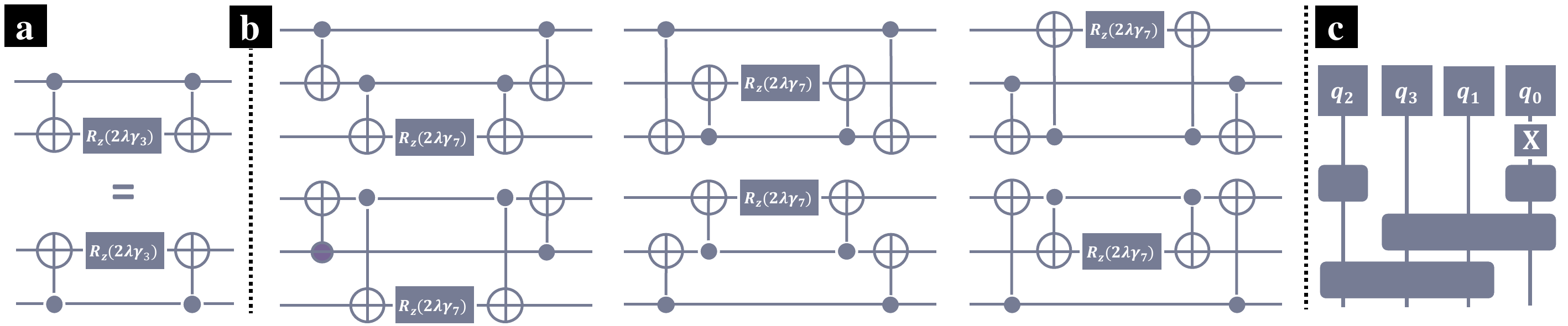}
    \caption{
    {\bf Connectivity in implementation of the $C^nR_y$ gates.}
    (a) (Upper) The implementation of $\exp(i\lambda\gamma_3\sigma_z\otimes\sigma_z)$.
    (Bottom) An equivalent implementation to the upper one, where we only swap the two qubits.
    (b) Six equivalent implementations of $\exp(i\lambda\gamma_7\sigma_z\otimes\sigma_z\otimes\sigma_z)$.
    (c) Schematic implementation of $U_{int}$ for the first part ERI of Helium.
    Four qubits $q_0, q_1, q_2, q_3$ are harnessed in the $U_{int}$, where $q_3$ represents the highest digit, and $q_0$ represents the lowest one.
    Here we apply the alternative $U_{int}^{alter}$ as defined in Eq.(\ref{eq_uint_alter}).
    }
    \label{figs_expo}
\end{figure}

Recall the implementation of ERI components.
In Fig.(\ref{Circuits estimating ERI}c,d), the implementation of $\exp(i\lambda\gamma_3\sigma_x\otimes\sigma_x)$ and $\exp(i\lambda\gamma_7\sigma_x\otimes\sigma_x\otimes\sigma_x)$ are presented.
The Hadamard gates change $\sigma_z$ basis into $\sigma_x$ basis, and the $zz$ interactions are introduced by CNOT gate pairs along with a single $R_z$ gate.
In Fig.(\ref{figs_expo}a), the upper is standard implementation of $\exp(i\lambda\gamma_3\sigma_z\otimes\sigma_z)$, which is same to the one in Fig.(\ref{Circuits estimating ERI}c).
Notice that the two qubits are symmetry in the $zz$ interaction.
Thereby the operation is equivalent to the original one when these two qubits are swapped, as shown in the bottom one in Fig.(\ref{figs_expo}a).
Similarly, there exist 6 equivalent implementation of the $zzz$ interactions.
In Fig.(\ref{figs_expo}b) the equivalent implementations of  $\exp(i\lambda\gamma_7\sigma_z\otimes\sigma_z\otimes\sigma_z)$ are presented, where the first implementation corresponds to the one in Fig.(\ref{Circuits estimating ERI}d).

Convenience is led by this feature of exponential products.
One typical example is the implementation of $U_{int}$ for the first part ERI of Helium, where there are only 4 non trivial components, corresponding to binary description $0000, 0101, 1010, 1111$.
As depicted in Fig.(\ref{figs_expo}c), four qubits $q_0, q_1, q_2, q_3$ are harnessed in the $U_{int}$, where $q_3$ represents the highest digit, and $q_0$ represents the lowest one.
Here we apply the alternative $U_{int}^{alter}$ as defined in Eq.(\ref{eq_uint_alter}).
At the very beginning, all qubits are initialized at ground state $|0\rangle$.
The $NOT$ gate on $q_0$ converts the quantum state to $|0001\rangle$.
Next, the solid square on $q_2$ indicates a $R_x$ gate acting on $q_2$, which contributes to $exp(i\lambda\sigma_x(q_2)\otimes\sigma_x(q_0))$, preparing the $0101$ term.
Meanwhile, there is a $Rx$ gate acting in $q_0$, corresponding to the $0000$ term.
Then consider the $1010$ term.
Recalling Eq.(\ref{eq_uint_alter}), the $1010$ term corresponds to the $zzz$ interaction in $exp(i\lambda\sigma_x(q_3)\otimes\sigma_x(q_1)\otimes\sigma_x(q_0))$, due to the existence of $NOT$ gate applied on $q_0$.
The long solid box that spans over $q_3, q_2, q_0$ indicates the implementation of the $zzz$ interaction.
Here $q_1$ is neighbor to both $q_3$ and $q_0$.
Consider the first implementation in Fig.(\ref{figs_expo}b), assign $q_3$ the first qubit, $q_1$ the second and $q_0$ the last, then there is only $CNOT$ gates acting between the neighbors.
On the other hand, the $1010$ term also corresponds to the $zzz$ interaction among $q_3$, $q_2$, $q_1$, as the long solid box (bottom) that spans over $q_3, q_2, q_1$ depicted in  Fig.(\ref{figs_expo}c).
Now consider the last implementation in Fig.(\ref{figs_expo}b), assign $q_3$ the first qubit, $q_2$ the second and $q_1$ the last, similarly, there is only $CNOT$ gates acting between the neighbors, say $q_1$ and $q_3$, $q_3$ and $q_2$.

To implement $CNOT$ gate between two qubits that are not directly connected on real devices, $SWAP$ gates are inevitable, which often causes extra noises and errors.
Thereby, we prefer the $CNOT$ gates applied on the connected neighbors.
Fortunately, under appropriate mapping as discussed in this subsection, the Pauli-X products can often be implemented with simple Hadamard gates, single qubit $R_z$ gate, and a sequence of $CNOT$ gates acting between the connected neighbors.

\subsection{Estimation of the denominators}
In the main article, operation $U^{(all)}_E$ is designed as shown in Eq.(\ref{eq_u_e_sqrt}).
For simplicity, hereafter we denote $P_y^x$ as the theoretical counts to find output state $|y\rangle$ with input state $|x\rangle$, where $x$ and $y$ are integers representing the state of $q'$(as the first digit in the binary form) and $q$.
In the main article, we focus on the simple Helium atom, and 4 $q$ qubits are involved in $U_{INT}$.
State $|x\rangle$ represents $|0\rangle_{q'}\otimes|(x)_{bin}\rangle_{q}$, whereas state $|x+16\rangle$ represents $|1\rangle_{q'}\otimes|(x)_{bin}\rangle_{q}$, where $(x)_{bin}$ indicates the binary form of integer $x$, and subscripts $q$ or $q'$ indicate the corresponding qubits.
Theoretically, we have
\begin{equation}
    P^x_{x+16} = \left|
    \frac{C_e^{sqrt}}{\epsilon_a+\epsilon_b-\epsilon_r-\epsilon_s}\cdot \right|M_{shots}
\end{equation}
and
\begin{equation}
    P^x_{x} = 
    \left(
    1-\left|
    \frac{C_e^{sqrt}}{\epsilon_a+\epsilon_b-\epsilon_r-\epsilon_s}
    \right|\right)\cdot M_{shots}
\end{equation}
where the total number of shots is denoted as $M_{shots}$.
Ideally, there is no other output except $|x\rangle$ and $|x=16\rangle$, ensuring that
\begin{equation}
    \left|\frac{C_e}{\epsilon_a+\epsilon_b-\epsilon_r-\epsilon_s}\right| 
    =\frac{P^x_{x+16}}{P_x^x + P^x_{x+16}}
\end{equation}

However, noise can not be ignored in real quantum devices.
Recalling the results as presented in the left column of Fig.(2c,d,e,f), unexpected states take considerable ratio in the $U_{INT}^{(all)}$ outputs.
In order to figure out the root of these unexpected patterns, we tested $U_{INT}^{(lite)}$, where only necessary operations are included, and the results are presented in the right column of Fig.(2c,d,e,f).
Similarly to the notations in the main article, $p_{x}^y$ is the count to find output $x$ with input $y$ with operation $U_E^{all}$, $\tilde{P}$ corresponds to the count under operation $U_E^{lite}$.

We notice that outputs of $U_{INT}^{(lite)}$ are always very close to theoretical prediction, except the $\tilde{P}^0$ terms. 
Thus the denominator terms can be approximated as
\begin{equation}
    \left|\frac{C_e}{\epsilon_a+\epsilon_b-\epsilon_r-\epsilon_s}\right| 
    =
    \frac{\tilde{P}_{x+16}^x}{\tilde{P}_x^x + \tilde{P}_{x+16}^x}
    ,\qquad
    x \neq 0
\end{equation}
Recalling that $\tilde{P}_0$ terms corresponding to input $|0\rangle\otimes|0000\rangle$ (or $|0\rangle$), where there is a $C^4R_y$ gate in $U_{INT}^{(lite)}$.
For all of the other terms, the $C^4R_y$ gate is eliminated, and the outputs are close to the theoretical prediction.
On the contrary, there is always a $C^4R_y$ gate in $U_{INT}^{(all)}$, no matter if it is working or idling.
Taking these above into account, we can figure out that most errors are owing to the existence of $C^4R_y$ gate.

Denote $U_{\Delta}$ as the difference between the operation $U^{(all)}_{E}$ in experiment and the ideal $U_E$, we have
\begin{equation}
    \frac{P_y^x}{P_x^x+P_{x+16}^x}
    =
    \left|
    \langle y|U_E|x\rangle
    \right|^2
\end{equation}
and
\begin{equation}
    \frac{p^x_{y}}{p_x^x + p^x_{x+16}}
    =
    \left|
    \langle y|U_E + U_{\Delta}|x\rangle
    \right|^2
\end{equation}
Rewrite $U_{\Delta}=\Delta U_E$, we have
\begin{equation}
    \begin{split}
        \frac{p^x_{y}}{p_x^x + p^x_{x+16}}
        &=
        \left|
        \langle y|({\mathbb{1}} + \Delta)U_E|x\rangle
        \right|^2
        \\
        &=
        \left|
        \langle y|({\mathbb{1}} + \Delta)
        \left(
        \sqrt{1-\left|\frac{C_e^{sqrt}}{\epsilon_a+\epsilon_b-\epsilon_r-\epsilon_s}\right|}
        |x\rangle
        +
        \sqrt{\left|\frac{C_e^{sqrt}}{\epsilon_a+\epsilon_b-\epsilon_r-\epsilon_s}\right|}
        |x+16\rangle
        \right)
        \right|^2
        \\
        &=
        \frac{1}{P_x^x + P^x_{x+16}}\left|
        \left(\delta_{xy}\sqrt{P_x^x}
        +\delta_{x+16, y}\sqrt{P_{x+16}^x}\right)
        +
        \sqrt{P_x^x}\langle y|\Delta|x\rangle
        +
        \sqrt{P_{x+16}^x}\langle y|\Delta|x+16\rangle
        \right|^2
    \end{split}
\end{equation}
In the approximation of the denominators, we focus on the outputs $|x\rangle$ and $|x+16\rangle$.
We have
\begin{equation}
    \begin{split}
        \frac{p^x_{x}}{p_x^x + p^x_{x+16}} &=
    \frac{1}{P_x^x + P^x_{x+16}}\left|
    \sqrt{P_x^x}
    +\sqrt{P_x^x}\langle x|\Delta|x\rangle
    +\sqrt{P_{x+16}^x}\langle x|\Delta|x+16\rangle
    \right|^2
    \\
    &=
    \frac{P_x^x}{P_x^x + P^x_{x+16}}\cdot
    \left|
    1+\langle x|\Delta|x\rangle
    +\sqrt{\frac{P_{x+16}^x}{P_{x}^x}}\langle x|\Delta|x+16\rangle
    \right|^2
    \end{split}
\end{equation}
Meanwhile,
\begin{equation}
    \begin{split}
        \frac{p^x_{x+16}}{p_x^x + p^x_{x+16}} &=
        \frac{1}{P_x^x + P^x_{x+16}}\left|\sqrt{P_{x+16}^x}
        +\sqrt{P_x^x}\langle x+16|\Delta|x\rangle
        +\sqrt{P_{x+16}^x}\langle x+16|\Delta|x+16\rangle
        \right|^2
        \\
        &=
        \frac{P_{x+16}^x}{P_x^x + P^x_{x+16}}\cdot
        \left|1+\langle x+16|\Delta|x+16\rangle
        +\sqrt{\frac{P_x^x}{P_{x+16}^x}}\langle x+16|\Delta|x\rangle
        \right|^2
    \end{split}
\end{equation}

Recalling the decomposition of the $C^4R_y$ gate as shown in Fig.(2a), we can notice that all of the four qubits in $q$ are equivalent.
Symmetry in the quantum circuit guarantees that $\langle x|\Delta|x\rangle$ should be a same value for $x=0,1,\cdots,15$, whereas $\langle x+16|\Delta|x+16\rangle$ should be another same value.
As most errors are on account of the pair of Toffoli gates pairs, $\Delta$ can hardly flip $q'$ without adding any other changes.
In other words, it is expected that 
$|\langle x|\Delta|x+16\rangle|\ll|\langle x|\Delta|x\rangle|$ 
and
$|\langle x+16|\Delta|x\rangle|\ll|\langle x+16|\Delta|x+16\rangle|$.

Therefore we have
\begin{equation}
    \frac{p^x_{x+16}}{p_x^x + p^x_{x+16}}=
    \frac{1}{15}\sum_{n=1}^{16}\left(1-
    \frac{\tilde{P}_{n+16}^n-p^n_{n+16}}{\tilde{P}_n^n + \tilde{P}_{n+16}^n}
    \right)\cdot
    \frac{P^x_{x+16}}{P_x^x + P^x_{x+16}}
\end{equation}
where we ignored the $\langle x+16|\Delta|x\rangle$ terms.
and the denominators terms are estimated as
\begin{equation}
    \left|\frac{C_e}{\epsilon_a+\epsilon_b-\epsilon_r-\epsilon_s}\right| 
    =
    \left(\frac{1}{15}\sum_{n=1}^{16}
    \frac{\tilde{P}_n^n+p^n_{n+16}}{\tilde{P}_n^n + \tilde{P}_{n+16}^n}
    \right)^{-1}\cdot
    \frac{p^x_{x+16}}{p_x^x + p^x_{x+16}}
\end{equation}
which is the approximation as shown in the main article.

\end{document}